\documentclass[12pt,a4paper]{article}

\usepackage{ifthen} 
\newboolean{pdflatex}
\setboolean{pdflatex}{true} 
\newboolean{articletitles}
\setboolean{articletitles}{true} 
\newboolean{uprightparticles}
\setboolean{uprightparticles}{false} 
\newboolean{inbibliography}
\setboolean{inbibliography}{false} 

\usepackage{microtype}
\usepackage{lineno}  
\usepackage{xspace} 

\usepackage{graphicx}  
\usepackage{color}
\usepackage{colortbl}
\graphicspath{{./figs/}} 

\usepackage{amsmath} 
\usepackage{amssymb}
\usepackage{amsfonts}
\usepackage{mathrsfs,bm,url,color}
\usepackage{upgreek} 

\newcommand*\patchAmsMathEnvironmentForLineno[1]{%
\expandafter\let\csname old#1\expandafter\endcsname\csname #1\endcsname
\expandafter\let\csname oldend#1\expandafter\endcsname\csname
end#1\endcsname
 \renewenvironment{#1}%
   {\linenomath\csname old#1\endcsname}%
   {\csname oldend#1\endcsname\endlinenomath}%
}
\newcommand*\patchBothAmsMathEnvironmentsForLineno[1]{%
  \patchAmsMathEnvironmentForLineno{#1}%
  \patchAmsMathEnvironmentForLineno{#1*}%
}
\AtBeginDocument{%
\patchBothAmsMathEnvironmentsForLineno{equation}%
\patchBothAmsMathEnvironmentsForLineno{align}%
\patchBothAmsMathEnvironmentsForLineno{flalign}%
\patchBothAmsMathEnvironmentsForLineno{alignat}%
\patchBothAmsMathEnvironmentsForLineno{gather}%
\patchBothAmsMathEnvironmentsForLineno{multline}%
}

\usepackage{hyperref}    
\usepackage[all]{hypcap} 

\def\ux85 {\mbox{UX85}\xspace}

\ifthenelse{\boolean{uprightparticles}}%
{

 \def\Ppi         {\ensuremath{\uppi}\xspace}

 \def\PDelta      {\ensuremath{\Delta}\xspace}                 
 \def\PXi      {\ensuremath{\Xi}\xspace}                 
 \def\PLambda      {\ensuremath{\Lambda}\xspace}                 
 \def\PSigma      {\ensuremath{\Sigma}\xspace}                 
 \def\POmega      {\ensuremath{\Omega}\xspace}                 
 \def\PUpsilon      {\ensuremath{\Upsilon}\xspace}                 
 
 \def\PB      {\ensuremath{\mathrm{B}}\xspace}                 
                  
 \def\PD      {\ensuremath{\mathrm{D}}\xspace}

 \def\PK      {\ensuremath{\mathrm{K}}\xspace}

 \def\Pi      {\ensuremath{\mathrm{i}}\xspace}

 \def\Ps      {\ensuremath{\mathrm{s}}\xspace}

}
{

 \def\Ppi         {\ensuremath{\pi}\xspace}

 \mathchardef\PDelta="7101
 \mathchardef\PXi="7104
 \mathchardef\PLambda="7103
 \mathchardef\PSigma="7106
 \mathchardef\POmega="710A
 \mathchardef\PUpsilon="7107
                  
 \def\PB      {\ensuremath{B}\xspace}                 
                  
 \def\PD      {\ensuremath{D}\xspace}

 \def\PK      {\ensuremath{K}\xspace}

 \def\Pi      {\ensuremath{i}\xspace}

 \def\Ps      {\ensuremath{s}\xspace}

}

\def\squark    {\ensuremath{\Ps}\xspace}

\def\pion  {\ensuremath{\Ppi}\xspace}
\def\piz   {\ensuremath{\pion^0}\xspace}

\def\pip   {\ensuremath{\pion^+}\xspace}
\def\pim   {\ensuremath{\pion^-}\xspace}

\def\pipm  {\ensuremath{\pion^\pm}\xspace}

\def\kaon  {\ensuremath{\PK}\xspace}
  \def\Kbar  {\kern 0.2em\overline{\kern -0.2em \PK}{}\xspace}

\def\Kz    {\ensuremath{\kaon^0}\xspace}
\def\Kzb   {\ensuremath{\Kbar^0}\xspace}
\def\KzKzb {\ensuremath{\Kz \kern -0.16em \Kzb}\xspace}
\def\Kp    {\ensuremath{\kaon^+}\xspace}
\def\Km    {\ensuremath{\kaon^-}\xspace}

\def\Kmp   {\ensuremath{\kaon^\mp}\xspace}
\def\KpKm  {\ensuremath{\Kp \kern -0.16em \Km}\xspace}
\def\KS    {\ensuremath{\kaon^0_{\rm\scriptscriptstyle S}}\xspace}

\def\Kstar   {\ensuremath{\kaon^*}\xspace}

\def\Dbar    {\kern 0.2em\overline{\kern -0.2em \PD}{}\xspace}
\def\D       {\ensuremath{\PD}\xspace}

\def\Dz      {\ensuremath{\D^0}\xspace}
\def\Dzb     {\ensuremath{\Dbar^0}\xspace}
\def\DzDzb   {\ensuremath{\Dz {\kern -0.16em \Dzb}}\xspace}
\def\Dp      {\ensuremath{\D^+}\xspace}
\def\Dm      {\ensuremath{\D^-}\xspace}

\def\DpDm    {\ensuremath{\Dp {\kern -0.16em \Dm}}\xspace}
\def\Dstar   {\ensuremath{\D^*}\xspace}

\def\B       {\ensuremath{\PB}\xspace}
\def\Bbar    {\ensuremath{\kern 0.18em\overline{\kern -0.18em \PB}{}}\xspace}

\def\Bz      {\ensuremath{\B^0}\xspace}
\def\Bzb     {\ensuremath{\Bbar^0}\xspace}

\def\Bs      {\ensuremath{\B^0_\squark}\xspace}
\def\Bsb     {\ensuremath{\Bbar^0_\squark}\xspace}

  \def\Y#1S{\ensuremath{\PUpsilon{(#1S)}}\xspace}

\def\Lbar {\ensuremath{\kern 0.1em\overline{\kern -0.1em\PLambda}}\xspace}

\def\to                 {\ensuremath{\rightarrow}\xspace}

\def\CP                {\ensuremath{C\!P}\xspace}

\def\AT#1     {\ensuremath{A_{\mathrm{T}}^{#1}}\xspace}           

\def\C#1      {\ensuremath{\mathcal{C}_{#1}}\xspace}                       
\def\Cp#1     {\ensuremath{\mathcal{C}_{#1}^{'}}\xspace}                    
\def\Ceff#1   {\ensuremath{\mathcal{C}_{#1}^{\mathrm{(eff)}}}\xspace}        
\def\Cpeff#1  {\ensuremath{\mathcal{C}_{#1}^{'\mathrm{(eff)}}}\xspace}       
\def\Ope#1    {\ensuremath{\mathcal{O}_{#1}}\xspace}                       
\def\Opep#1   {\ensuremath{\mathcal{O}_{#1}^{'}}\xspace}                    

\def\ycp        {\ensuremath{y_{\CP}}\xspace}

\newcommand{\tev}{\ensuremath{\mathrm{\,Te\kern -0.1em V}}\xspace}
\newcommand{\gev}{\ensuremath{\mathrm{\,Ge\kern -0.1em V}}\xspace}
\newcommand{\mev}{\ensuremath{\mathrm{\,Me\kern -0.1em V}}\xspace}
\newcommand{\kev}{\ensuremath{\mathrm{\,ke\kern -0.1em V}}\xspace}
\newcommand{\ev}{\ensuremath{\mathrm{\,e\kern -0.1em V}}\xspace}
\newcommand{\gevc}{\ensuremath{{\mathrm{\,Ge\kern -0.1em V\!/}c}}\xspace}
\newcommand{\mevc}{\ensuremath{{\mathrm{\,Me\kern -0.1em V\!/}c}}\xspace}
\newcommand{\gevcc}{\ensuremath{{\mathrm{\,Ge\kern -0.1em V\!/}c^2}}\xspace}
\newcommand{\gevgevcccc}{\ensuremath{{\mathrm{\,Ge\kern -0.1em V^2\!/}c^4}}\xspace}
\newcommand{\mevcc}{\ensuremath{{\mathrm{\,Me\kern -0.1em V\!/}c^2}}\xspace}

\def\invfb   {\ensuremath{\mbox{\,fb}^{-1}}\xspace}

\def\gsim{{~\raise.15em\hbox{$>$}\kern-.85em
          \lower.35em\hbox{$\sim$}~}\xspace}
\def\lsim{{~\raise.15em\hbox{$<$}\kern-.85em
          \lower.35em\hbox{$\sim$}~}\xspace}

\def\tell1  {TELL1\xspace}
\def\ukl1   {UKL1\xspace}

\newcommand{\eg}{\mbox{\itshape e.g.}\xspace}

\usepackage{cite} 
\usepackage{mciteplus}
\usepackage{overpic}
\usepackage{subfigure}

\usepackage[noabbrev,capitalise]{cleveref}

\usepackage{floatrow}

\newfloatcommand{capbtabbox}{table}[][\FBwidth]

\usepackage{caption}
\usepackage{longtable}
\usepackage{ifpdf}

\newcommand{\re}[2][()] {\ifthenelse{\equal{#1}{()}}{{\ensuremath{{\rm \, Re}}\left(#2\right)}}
                                                    {{\ensuremath{{\rm \, Re}}\left[#2\right]}}}
\newcommand{\im}[2][()] {\ifthenelse{\equal{#1}{()}}{{\ensuremath{{\rm \, Im}}\left(#2\right)}}
                                                    {{\ensuremath{{\rm \, Im}}\left[#2\right]}}}

\definecolor{orange}{rgb}{1,0.5,0}

\usepackage{booktabs}
\usepackage{rotating}
\usepackage{multirow}

\newcommand{\massgevsq}{\mbox{\ensuremath{\gev^2\!/c^4}}\xspace}

\newcommand{\kspp}{\ensuremath{\KS\pip\pim}\xspace}
\newcommand{\Dkspp}{\ensuremath{\Dz\to\KS\pip\pim}\xspace}

\def\Fbar{\kern 0.2em\overline{\kern -0.2em F}{}\xspace}
\def\Xbar{\kern 0.2em\overline{\kern -0.2em X}{}\xspace}
\def\Nbar{\kern 0.2em\overline{\kern -0.2em N}{}\xspace}
\def\Rbar{\kern 0.2em\overline{\kern -0.2em R}{}\xspace}

\newcommand{\zcp}{\ensuremath{z_{\CP}}\xspace}
\newcommand{\xcp}{\ensuremath{x_{\CP}}\xspace}
\renewcommand{\ycp}{\ensuremath{y_{\CP}}\xspace}
\newcommand{\deltaz}{\ensuremath{\Delta z}\xspace}
\newcommand{\deltax}{\ensuremath{\Delta x}\xspace}
\newcommand{\deltay}{\ensuremath{\Delta y}\xspace}

\newcommand{\0}{\phantom{0}}

\begin{document}
\renewcommand{\thefootnote}{\fnsymbol{footnote}}
\setcounter{footnote}{1}
\begin{titlepage}


\vspace*{1.5cm}

{\bf\boldmath\huge
\begin{center}
Novel method for measuring charm-mixing parameters using multibody decays
\end{center}
}

\vspace*{0.5cm}

\begin{center}
A.~Di~Canto$^1$,
J.~Garra~Tic\'{o}$^2$,
T.~Gershon$^3$,
N.~Jurik$^4$,
M.~Martinelli$^1$,
T.~Pila\v{r}$^3$,
S.~Stahl$^1$,
D.~Tonelli$^5$
\bigskip\\
{\it\footnotesize
$^1$European Organization for Nuclear Research (CERN), Geneva, Switzerland\\
$^2$Cavendish Laboratory, University of Cambridge, Cambridge, United Kingdom\\
$^3$Department of Physics, University of Warwick, Coventry, United Kingdom\\
$^4$Department of Physics, University of Oxford, Oxford, United Kingdom\\
$^5$INFN Sezione di Trieste, Trieste, Italy}
\end{center}

\vspace{\fill}

\begin{abstract}
\noindent
We propose a novel method to measure flavor-oscillations and charge-parity (\CP) violation in charm mixing. The approach applies to multibody charm decays, such as \mbox{\Dkspp}, and avoids the need for a fit of the decay amplitudes while suppressing biases due to nonuniform signal-reconstruction efficiencies as functions of phase space and decay time. Data are partitioned in decay-time and Dalitz-plot regions (bins). The Dalitz-plot bins are symmetric with respect to the principal bisector and chosen to ensure nearly constant values of the strong-interaction phases in each. The ratios of signal yields observed in each symmetric bin pair are fit as functions of decay time, using independent auxiliary measurements of the strong-interaction phases as constraints, to determine the relevant physics parameters. Simulation shows a 35\% improvement in sensitivity to the normalized charm-eigenstate mass difference with respect to existing model-independent methods. In addition, we introduce a parametrization of oscillation and \CP-violation effects in charm mixing that has attractive statistical properties and may find wider applicability. 
\end{abstract}

\vspace*{1.5cm}

\begin{center}
  Published in \href{https://doi.org/10.1103/PhysRevD.99.012007}{Phys.\ Rev.\ {\bf D99} (2019) 012007}
\end{center}

\vspace{\fill}

\end{titlepage}

\pagestyle{empty}  


\renewcommand{\thefootnote}{\arabic{footnote}}
\setcounter{footnote}{0}
\tableofcontents
\cleardoublepage
\pagestyle{plain} 
\setcounter{page}{1}
\pagenumbering{arabic}


\graphicspath{{figs/}}
\allowdisplaybreaks

\section{Introduction}

The noncoincidence of mass and flavor eigenstates of neutral flavored mesons results in flavor oscillations, which are meson-antimeson transitions that follow an oscillating pattern as a function of time. Flavor oscillations are sensitive probes for non-standard-model physics since virtual massive particles can contribute to the amplitude, possibly enhancing the average oscillation rate or the difference between rates of mesons and those of their respective antimesons. Indeed, the study of flavor oscillations has long been established as a powerful instrument to uncover, or constrain, possible dynamics not described by the standard model.

Oscillations are typically characterized by the dimensionless mixing parameters \mbox{$x \equiv \Delta m /\Gamma$} and \mbox{$y \equiv \Delta \Gamma/ 2\Gamma$}, where $\Delta m$ ($\Delta\Gamma$) is the difference between the masses (decay widths) of the neutral-meson eigenstates, and $\Gamma$ is the average decay width~\cite{pdg}. Oscillations were first observed in the \Kz--\Kzb system in 1956~\cite{Lande:1956pf}, then established in the \Bz--\Bzb system in 1987~\cite{Albrecht:1987dr}, and in the \Bs--\Bsb system in 2006~\cite{Abulencia:2006ze}. Oscillation parameters for all these mesons are known precisely, except for the width-difference of \Bz mesons~\cite{pdg}. The first evidence for \Dz--\Dzb oscillations was reported in 2007~\cite{Aubert:2007wf,Staric:2007dt} and the first single-experiment observation in 2012~\cite{LHCb-PAPER-2012-038}. However, the underlying charm-mixing parameters still have significant uncertainties. Recent global combinations yield $x = (4.6\,^{+\,1.2}_{-\,1.3})\times10^{-3}$ and $y = (6.2\pm0.7)\times10^{-3}$, assuming charge-parity (\CP) symmetry of doubly Cabibbo-suppressed decay amplitudes~\cite{hfag}. While the global knowledge of $y$ is rather precise, less is known about $x$, which has not even been conclusively shown to differ from zero. Improving the knowledge of $x$ is especially critical since sensitivity to the small phase $\phi$ that describes \CP violation in the interference between mixing and decay relies predominantly on observables proportional to $x\sin\phi$.

The most direct experimental access to the charm-mixing parameters is offered by the analysis of self-conjugate multibody decays, such as \Dkspp (inclusion of charge-conjugate processes is implied unless stated otherwise). A joint fit of the Dalitz-plot and decay-time distributions allows for the determination of a \Dz component growing as a function of decay time in a sample of candidates produced as \Dzb mesons, and vice versa. This approach is challenging as it requires fitting the decay-time evolution of signal decays across the Dalitz plot with an accurate amplitude model, accounting for efficiency and resolution effects, and backgrounds components~\cite{Asner:2005sz,Peng:2014oda,delAmoSanchez:2010xz}.

With the large samples of \Dkspp decays expected at the LHCb and Belle~II experiments~\cite{Bediaga:2018lhg,Kou:2018nap}, the systematic uncertainties due to knowledge of the amplitude model are likely to limit the final precision on the mixing parameters.  Approaches that obviate the need for an amplitude analysis of the Dalitz-plot distribution have been proposed to mitigate this issue~\cite{Bondar:2010qs,Thomas:2012qf}. These build on ideas developed to measure the CKM angle $\gamma$ from \mbox{$B^-\to\D(\to \KS\pip\pim)K^-$} decays, known as the GGSZ method~\cite{Giri:2003ty,Bondar:2005ki,Bondar:2008hh}. By partitioning the Dalitz plot into bins, the need for an explicit amplitude model is avoided, and the decay-time distribution depends on a small number of coefficients that encode relevant information about the decay, in addition to the mixing parameters. At hadron-collider experiments, however, such model-independent methods still face a significant challenge. Stringent online event-selection criteria are imposed on charged-particle momenta and displacements from the primary interaction space-point to suppress the prevailing backgrounds from light-quark production. Modeling the resulting biases on signal decay-time and Dalitz-plot distributions increases the complexity of the analyses introducing further sources of systematic uncertainty that may offset the intended advantages~\cite{LHCb-PAPER-2015-042}.

We propose a novel approach for measuring parameters of oscillation and \CP violation in charm mixing using \Dkspp, or other multibody neutral-charm decays, that requires neither an amplitude analysis of the Dalitz-plot distribution nor an accurate modeling of the efficiency variations as functions of decay time and Dalitz-plot position. The sample of \Dkspp decays is divided into subsamples according to initial \Dz meson flavor, location on the Dalitz plot, and decay time. Ratios of decay yields observed in regions (``bins'') of the Dalitz plot that are symmetric about its bisector are constructed as functions of decay time. These functions depend on the known hadronic parameters, dependent on Dalitz-plot bin, that enter the GGSZ method to determine $\gamma$~\cite{Libby:2010nu}. The mixing parameters are obtained from a least-squares fit of the decay-time-dependent ratios, jointly for mesons produced as \Dz and \Dzb, in which external information on the hadronic parameters is used as a constraint. Any significant \CP-violating effect in oscillations of \Dz and \Dzb mesons is observed as a difference in the ratios between the samples of mesons produced in the \Dz and \Dzb states. We dub this approach the ``bin-flip method''.

In \cref{sec:method} we develop the formalism of the method; in \cref{sec:cleo-inputs} we discuss the Dalitz-plot partition and external inputs needed; in \cref{sec:sensitivity} we evaluate the sensitivity using simulated samples and discuss instrumental effects such as those due to resolutions and nonuniform reconstruction efficiencies; in \cref{sec:impact} we quantify the impact of the method on the knowledge of charm-mixing phenomenology to finally conclude in \cref{sec:conclusions}.

\section{The bin-flip method}\label{sec:method}
Mass eigenstates of neutral charm mesons are expressed as $|D_{1,2}\rangle = p|\Dz\rangle  \pm  q |\Dzb\rangle$ in terms of flavor eigenstates, where $p$ and $q$ are complex parameters satisfying $|q|^2+|p|^2=1$. In the limit of \CP symmetry ($q=p$), we define $D_{1(2)}$ to be the \CP-even (odd) eigenstate, and the mixing parameters as $x = (m_1 - m_2)/\Gamma$ and $y =(\Gamma_1 - \Gamma_2)/(2\Gamma)$, where $\Gamma=(\Gamma_1+\Gamma_2)/2$ is the average decay-width, following Refs.~\cite{pdg,hfag}. We specialize the discussion of the method to the \Dkspp decays, since we anticipate that it will have a strong impact when used with this mode, but the formalism can be adapted to other multibody decays.

We parametrize the \Dkspp three-body decay dynamics with two two-body masses following the Dalitz formalism~\cite{Dalitz:1953cp,Fabri:1954zz}. We use the following flavor-dependent definition of squared invariant masses:
\begin{equation}
m_\pm^2 \equiv \left\{\begin{aligned}
m^2(\KS\pi^\pm)\quad &\text{for}\ \Dz \to\kspp\ \text{decays}\\
m^2(\KS\pi^\mp)\quad &\text{for}\ \Dzb\to\kspp\ \text{decays}\\
\end{aligned}\right.,
\end{equation}
which simplifies the simultaneous treatment of \Dz and \Dzb decays.

We indicate with $A_f(m_+^2,m_-^2)$ and $\bar{A}_f(m_+^2,m_-^2)$ the amplitudes for mesons produced as \Dz and \Dzb, respectively, and decaying to the final state $f=\kspp$ at the generic point $(m_+^2,m_-^2)$ of the Dalitz plane. If \CP symmetry is conserved in the decay, the relation \mbox{$A_f(m_+^2,m_-^2)=\bar{A}_f(m_+^2,m_-^2)$} holds. The decay rates of neutral \D mesons tagged in the flavor eigenstates \Dz and \Dzb at time $t=0$ evolve in time as
\begin{align}
\left|{T}_f(m_+^2,m_-^2;t)\right|^2 &= \left|{A}_f(m_+^2,m_-^2)\, g_+(t) + \bar{A}_f(m_-^2,m_+^2)\, \frac{q}{p}\, g_-(t)\right|^2\quad\text{and} \label{eq:rate-D0}\\
\left|\overline{T}_{\!f}(m_+^2,m_-^2;t)\right|^2 &= \left|\bar{A}_f(m_+^2,m_-^2)\, g_+(t) + {A}_f(m_-^2,m_+^2)\, \frac{p}{q}\, g_-(t)\right|^2,\label{eq:rate-D0bar}
\end{align}
where $g_\pm(t) = \theta(t)e^{-imt}e^{-t/2}~^{\cosh}_{\sinh}(zt/2)$, $t$ is the decay time in units of \Dz lifetime $\tau=1/\Gamma$, $m=(m_1+m_2)/2$ is the average mass of neutral \D mesons, $\theta$ is the Heaviside function, and $z$ equals $-(y+ix)$.

We divide the Dalitz plane into two sets of $n$ bins each, symmetric about its principal bisector $m_+^2 = m_-^2$. Bins are labeled with the index $\pm b$, where $b=1,...,n$. Positive indices refer to bins in the (lower) $m_+^2 > m_-^2$ region, where Cabibbo-favored \mbox{$\Dz\to\Kstar(892)^-\pi^+$} decays dominate the amplitude; negative indices refer to their symmetric counterparts in the (upper) $m_+^2 < m_-^2$ region. 

As oscillations develop as a function of time, the relative variations of intensities between pairs of bins change depending on the mixing parameters and relevant charm-decay hadronic parameters. The expressions for the event yields integrated over each Dalitz-plot bin $b$ are 
\begin{align}
N_b(t) &= \int_b dm_+^2dm_-^2 \left| T_f(m_+^2,m_-^2;t) \right|^2\nonumber\\
  &= F_b \left| g_+(t) \right|^2 +
      \left| \frac{q}{p} \right|^2 \Fbar_{-b} \left| g_-(t) \right|^2 +
      2 \sqrt{\Fbar_{-b} F_b} \re[]{ \frac{q}{p} X_{b} \, g_+^\star(t) g_-(t) }\quad\text{and}\label{eq:nb1}\\
\Nbar_b(t) &= \int_b dm_+^2dm_-^2 \left| \overline{T}_{\!f}(m_+^2,m_-^2;t) \right|^2\nonumber\\
 &= \Fbar_b \left| g_+(t) \right|^2 +
     \left| \frac{p}{q} \right|^2 F_{-b} \left| g_-(t) \right|^2 +
     2 \sqrt{F_{-b} \Fbar_b} \re[]{ \frac{p}{q} \bar{X}_{b} \, g_+^\star(t) g_-(t) },\label{eq:nb2}
\end{align}
where the following definitions are introduced:
\begin{gather}
{F}_b \equiv \int_b dm_+^2dm_-^2 \left|A_f(m_+^2,m_-^2)\right|^2,\quad
\Fbar_b \equiv \int_b dm_+^2dm_-^2 \left| \bar{A}_f(m_+^2,m_-^2)\right|^2,\\
X_b \equiv \frac{1}{\sqrt{F_b \Fbar_{-b}}} \int_b \! dm_+^2dm_-^2\,A_f^\star(m_+^2,m_-^2)\bar{A}_f(m_-^2,m_+^2), \label{eq:Xb}
\end{gather}
and $\Xbar_{b}$ is defined similarly as in \cref{eq:Xb} with $A_f \leftrightarrow \bar{A}_f$ and $F_b \leftrightarrow \Fbar_{b}$. Here, $F_b$ and $\Fbar_b$ are event yields in the Dalitz bin $b$ at $t=0$. The hadronic parameter of the interference term $X_b$, with (by definition) $X_{-b} = X_{b}^\star$ and $\left|X_b\right| \leq 1$, is related to the strong-interaction phase difference, $\Delta\delta$, and to the weak-interaction phase difference, $\varphi$, between $A_f(m_+^2,m_-^2)$ and $\bar{A}_f(m_-^2,m_+^2)$ averaged over bin $b$. In the limit of \CP-conserving decay amplitudes, $A_f = \bar{A}_f$ so $\Fbar_b = F_{b}$, $\varphi=0$, and $X_b=\Xbar_b$ hold. Hence, the real and imaginary parts of the coefficients \mbox{$X_b\equiv c_b-is_b$} are 
\begin{align}
c_b &\equiv \frac{1}{\sqrt{F_bF_{-b}}}\int_b dm_+^2dm_-^2\,\left|A_f (m^{2}_{+},m^{2}_{-})\right|\,\left|A_f(m^{2}_{-},m^{2}_{+})\right|\cos[\Delta\delta(m^{2}_{+},m^{2}_{-})]\quad\text{and}\\
s_b &\equiv \frac{1}{\sqrt{F_bF_{-b}}}\int_b dm_+^2dm_-^2\,\left|A_f (m^{2}_{+},m^{2}_{-})\right|\,\left|A_f(m^{2}_{-},m^{2}_{+})\right|\sin[\Delta\delta(m^{2}_{+},m^{2}_{-})],
\end{align}
where $\Delta\delta(m_{+}^2,m_{-}^2) = \delta(m_{+}^2,m_{-}^2)-\delta(m_{-}^2,m_{+}^2)$ and $\delta(m_{+}^2,m_{-}^2)$ is the phase of $A_f(m_{+}^2,m_{-}^2)$. Constraining the hadronic parameters $c_b$ and $s_b$ from independent external measurements offers access to the mixing parameters. 

If the probability $\epsilon(m_+^2,m_-^2)$ to select and reconstruct the decays is nonuniform across the Dalitz plane, the parameters $F_b$ and $(c_b,s_b)$ become 
\begin{gather}
\tilde{F}_{b} \equiv \int_b dm_+^2dm_-^2\,\epsilon(m_+^2,m_-^2)\,\left|A_f(m_+^2,m_-^2)\right|^2,\quad\text{and} \label{eq:Ftilde} \\
\tilde{c}_b \equiv \frac{1}{\sqrt{\tilde{F}_b\tilde{F}_{-b}}}\!\int_b\!dm_+^2dm_-^2\,\epsilon(m_+^2,m_-^2)\!\left|A_f (m^{2}_{+},m^{2}_{-})\right|\!\left|A_f(m^{2}_{-},m^{2}_{+})\right|\cos[\Delta\delta(m^{2}_{+},m^{2}_{-})], \label{eq:cbtilde}\\
\tilde{s}_b \equiv \frac{1}{\sqrt{\tilde{F}_b\tilde{F}_{-b}}}\!\int_b\!dm_+^2dm_-^2\,\epsilon(m_+^2,m_-^2)\!\left|A_f (m^{2}_{+},m^{2}_{-})\right|\!\left|A_f(m^{2}_{-},m^{2}_{+})\right|\sin[\Delta\delta(m^{2}_{+},m^{2}_{-})], \label{eq:sbtilde}
\end{gather}
respectively. It is important that efficiency-induced biases on $(c_b,s_b)$ are kept small, since these values will be constrained to externally measured values. This can be achieved by designing selection strategies aimed at minimizing biases on the Dalitz-plot distribution, when possible. Otherwise, efficiencies that are nonuniform but still symmetric across the Dalitz-plot bisector are expected to induce reduced  biases on $(c_b,s_b)$. In addition, appropriate choices of binning schemes may also mitigate the biases on $(c_b,s_b)$ induced by efficiency variations. For example, Dalitz bins defined such that $\Delta\delta(m^2_+,m^2_-)$ is nearly constant within each bin are expected to reduce the effect of the nonuniformities of the efficiency on $(c_b,s_b)$. We neglect the effect of efficiency variations as functions of Dalitz-plot position in the discussion of the method below, and discuss the possible biases in realistic experimental situations in \cref{sec:sensitivity:detectoreffects}.

\newcommand{\gpsq}{\displaystyle 1 + \frac{1}{4}\, \langle t^2\rangle_j\, \re{z^2}}
\newcommand{\gmsq}{\displaystyle \frac{1}{4}\, \langle t^2\rangle_j\, \left| z \right|^2}
\newcommand{\gpgm}{\displaystyle \frac{1}{2}\, \langle t\rangle_j\, z}
For small mixing parameters ($|z|t\ll1$), the following approximations hold:
\begin{align}
\left| g_+(t)\right|^2 &\approx e^{-t} + \frac{1}{4}\, e^{-t}\, t^2\, \re{z^2} + \mathcal{O}( z^4 ), \\
\left| g_-(t)\right|^2 &\approx \frac{1}{4}\, e^{-t}\, t^2\, |z|^2 + \mathcal{O}( z^4 ),\quad\text{and} \\
  g_+^\star(t) g_-(t)  &\approx \frac{1}{2}\, e^{-t}\, t\, z + \mathcal{O}( z^3 ).
\end{align}
Terms of $\mathcal{O}( z^3 )$ or higher can be neglected, so that integration of the above expressions over decay-time bin $j$ yields
\begin{align}
\int_j dt\, \left| g_+(t) \right|^2 &\approx n_j \left[\gpsq\right], \\
\int_j dt\, \left| g_-(t) \right|^2 &\approx n_j \gmsq,\quad\text{and} \\
\int_j dt\, g_+^\star(t) g_-(t)     &\approx n_j \gpgm,
\end{align}
where $\langle...\rangle_j$ denotes the average over the exponential distribution in the decay-time bin $j$, and $n_j$ is a normalization constant that cancels in ratios and is omitted in what follows. 

If the probability $\epsilon(t)$ to select and reconstruct the decays is nonuniform as a function of decay time within bin $j$, the average is performed over the observed decay-time distribution of mesons that did not undergo oscillation, $\epsilon(t) e^{-t}$. An advantage of the bin-flip method is that the dependence of results on $\epsilon(t)$ is minimal.

In the limit of \CP-conserving decay amplitudes, the decay yields  in Dalitz bin $b$ and decay-time bin $j$ of charm mesons originally produced in the \Dz or \Dzb flavor states are, respectively,
\begin{align}
N_{bj} &= \int_j dt N_b(t)\nonumber\\
 &\approx F_b \left[\gpsq\right] +
          \gmsq\left| \frac{q}{p} \right|^2 F_{-b} +
          \langle t\rangle_j \sqrt{F_{-b} F_b} \re{ \frac{q}{p} X_{b} z },\\
\Nbar_{bj} &= \int_j dt \overline{N}_b(t)\nonumber\\
 &\approx F_b \left[\gpsq\right] +
          \gmsq \left| \frac{p}{q} \right|^2 F_{-b} +
          \langle t\rangle_j \sqrt{F_{-b}F_b} \re{ \frac{p}{q} X_{b} z}.
\end{align}
For each decay-time bin $j$, the ratios between the decay yield in Dalitz bin $-b$ and Dalitz bin $b$, for mesons originally produced as \Dz or \Dzb are, respectively,
\begin{alignat}{3}
R_{bj} &= \frac{N_{-bj}}{N_{bj}}
 &&\approx \frac{r_b \left[\gpsq\right] + \gmsq\left|\frac{q}{p}\right|^2 + \langle t\rangle_j\sqrt{r_b}\re{X_b^\star\frac{q}{p}\,z}}{\gpsq+\gmsq r_b\left|\frac{q}{p}\right|^2 + \langle t\rangle_j\sqrt{r_b}\re{X_b\frac{q}{p}\,z}},\\
\Rbar_{bj} &= \frac{\Nbar_{-bj}}{\Nbar_{bj}}
 &&\approx \frac{r_b \left[\gpsq\right] + \gmsq\left|\frac{p}{q}\right|^2 + \langle t\rangle_j\sqrt{r_b} \re{X_b^\star\frac{p}{q}\,z}}{\gpsq+\gmsq r_b\left|\frac{p}{q}\right|^2 + \langle t\rangle_j\sqrt{r_b} \re{X_b\frac{p}{q}\,z}},
\end{alignat}
where $r_b = F_{-b}/F_b$.

\newcommand{\gpsqcp}{\displaystyle 1 + \frac{1}{4}\,\langle t^2\rangle_j\, \re{\zcp^2-\deltaz^2}}
\newcommand{\gmsqcpp}{\displaystyle\frac{1}{4}\,\langle t^2\rangle_j\, \left|\zcp+\deltaz\right|^2}
\newcommand{\gmsqcpm}{\displaystyle\frac{1}{4}\,\langle t^2\rangle_j\, \left|\zcp-\deltaz\right|^2}

The bin-flip approach consists in performing a joint fit of the $R_{bj}$ and $\Rbar_{bj}$ ratios to determine the oscillation and \CP-violation parameters in charm mixing, by constraining the coefficients $X_b$ from external measurements (\cref{sec:cleo-inputs}). Conceptually, this is akin to performing the wrong-sign-to-right-sign analysis of $\Dz\to K^{\mp}\pi^{\pm}$ decays~\cite{PhysRevLett.60.1239} simultaneously in specially chosen subsets of events, the Dalitz-plot bins, for which hadronic parameters are known. Unlike $\Dz\to K^{\mp}\pi^{\pm}$ decays, \Dkspp decays also proceed through amplitudes with defined \CP eigenvalues, with important consequences on the sensitivity of method, as discussed in \cref{sec:sensitivity}.

In practice, to avoid instabilities of the fit due to $(q/p)^{\pm1}z$ terms, where sensitivity to $q/p$ degrades if $\left| z \right| \approx 0$, we parametrize the ratios using $\zcp$ and $\deltaz$, defined by
\begin{equation}
\zcp\pm\deltaz\equiv\left(q/p\right)^{\pm1}z.
\end{equation}
With this definition,
\begin{equation}
z^2 = \left( \zcp + \deltaz \right) \left( \zcp - \deltaz \right) = \zcp^2 - \deltaz^2,\qquad\left( \frac{q}{p} \right)^{\!\!2} = \frac{ \zcp + \deltaz }{ \zcp - \deltaz },
\end{equation}
and the ratios become 
\begin{align}
R_{bj} &\approx \frac{r_b\left[\gpsqcp\right] + \gmsqcpp + \sqrt{r_b} \langle t\rangle_j \re[]{X_b^\star(\zcp+\deltaz)}}{\gpsqcp+r_b\,\gmsqcpp + \sqrt{r_b} \langle t\rangle_j \re[]{X_b(\zcp+\deltaz)}},
\label{eq:bin-flip-ratio-D0}\\
\Rbar_{bj} &\approx \frac{r_b\left[\gpsqcp\right] + \gmsqcpm + \sqrt{r_b} \langle t\rangle_j \re[]{X_b^\star(\zcp-\deltaz)}}{\gpsqcp+r_b\,\gmsqcpm + \sqrt{r_b} \langle t\rangle_j \re[]{X_b(\zcp-\deltaz)}}.
\label{eq:bin-flip-ratio-D0bar}
\end{align}
Using the customary convention for the charm-mixing \CP-violation phase $\phi\equiv\arg(q\bar{A}_f/pA_f)\approx\arg(q/p)$, which assumes the absence of any final-state-dependent weak-interaction phase between decay amplitudes (consistent with the limit of \CP-symmetric decay amplitudes), the interpretation of $\zcp$ and $\deltaz$ in terms of the usual mixing parameters becomes straightforward,
\begin{alignat}{3}
\xcp &= -\im{\zcp} &&= \frac{1}{2}\left[x\cos\phi\left(\left|\frac{q}{p}\right|+\left|\frac{p}{q}\right|\right)+y\sin\phi\left(\left|\frac{q}{p}\right|-\left|\frac{p}{q}\right|\right)\right], \label{eq:xcp-def}\\
\deltax &= -\im{\deltaz} &&= \frac{1}{2}\left[x\cos\phi\left(\left|\frac{q}{p}\right|-\left|\frac{p}{q}\right|\right)+y\sin\phi\left(\left|\frac{q}{p}\right|+\left|\frac{p}{q}\right|\right)\right], \label{eq:dx-def}\\
\ycp &= -\re{\zcp} &&= \frac{1}{2}\left[y\cos\phi\left(\left|\frac{q}{p}\right|+\left|\frac{p}{q}\right|\right)-x\sin\phi\left(\left|\frac{q}{p}\right|-\left|\frac{p}{q}\right|\right)\right], \label{eq:ycp-def}\\
\deltay &= -\re{\deltaz} &&= \frac{1}{2}\left[y\cos\phi\left(\left|\frac{q}{p}\right|-\left|\frac{p}{q}\right|\right)-x\sin\phi\left(\left|\frac{q}{p}\right|+\left|\frac{p}{q}\right|\right)\right]. \label{eq:dy-def}
\end{alignat}
Conservation of \CP symmetry in mixing ($|q/p|=1$) and in the interference of mixing and decay ($\phi = 0$) implies $\xcp=x$, $\ycp=y$, and $\deltax=\deltay=0$. The observables \deltay, frequently denoted as $A_\Gamma$, and \ycp are well known. The introduction of \xcp and \deltax allows for a conveniently symmetric notation and yields parameters with statistical properties optimally suited for use in measurements and combinations of results, as discussed in \cref{app:cpvparametrization}.
\section{Dalitz-plot partition and strong-interaction phase inputs\label{sec:cleo-inputs}}

Various Dalitz-plot binning schemes were developed by the CLEO collaboration for measuring the coefficients \mbox{$X_b\equiv c_b-is_b$}~\cite{Libby:2010nu}. These include schemes aimed at minimizing the variations of the strong-interaction phase differences across each bin, as well as alternatives explicitly designed to optimize the GGSZ sensitivity to $\gamma$. For the bin-flip method we propose to use the ``iso-$\Delta\delta$'' scheme with $n=8$ bins defined in each Dalitz semispace such that
\begin{equation}
2\pi(b - 3/2)/n < \Delta\delta(m_{+}^2,m_{-}^2) < 2\pi(b - 1/2 )/n,\quad b=1,...,n,
\end{equation}
where the variation of $\Delta\delta(m_{+}^2,m_{-}^2)$ over the Dalitz plane is evaluated using the ``BaBar 2008'' amplitude model~\cite{Aubert:2008bd}. Because this scheme keeps the strong-interaction phase difference approximately constant in each Dalitz-plot bin, biases due to nonuniform efficiencies are reduced. A dedicated binning optimization for the bin-flip method may lead to improved sensitivity, but this is not pursued here, since we intend to rely on existing measurements of the hadronic parameters to demonstrate quantitatively the performance of the method.

\begin{figure}[ht]
\centering
\includegraphics[width=0.6\textwidth]{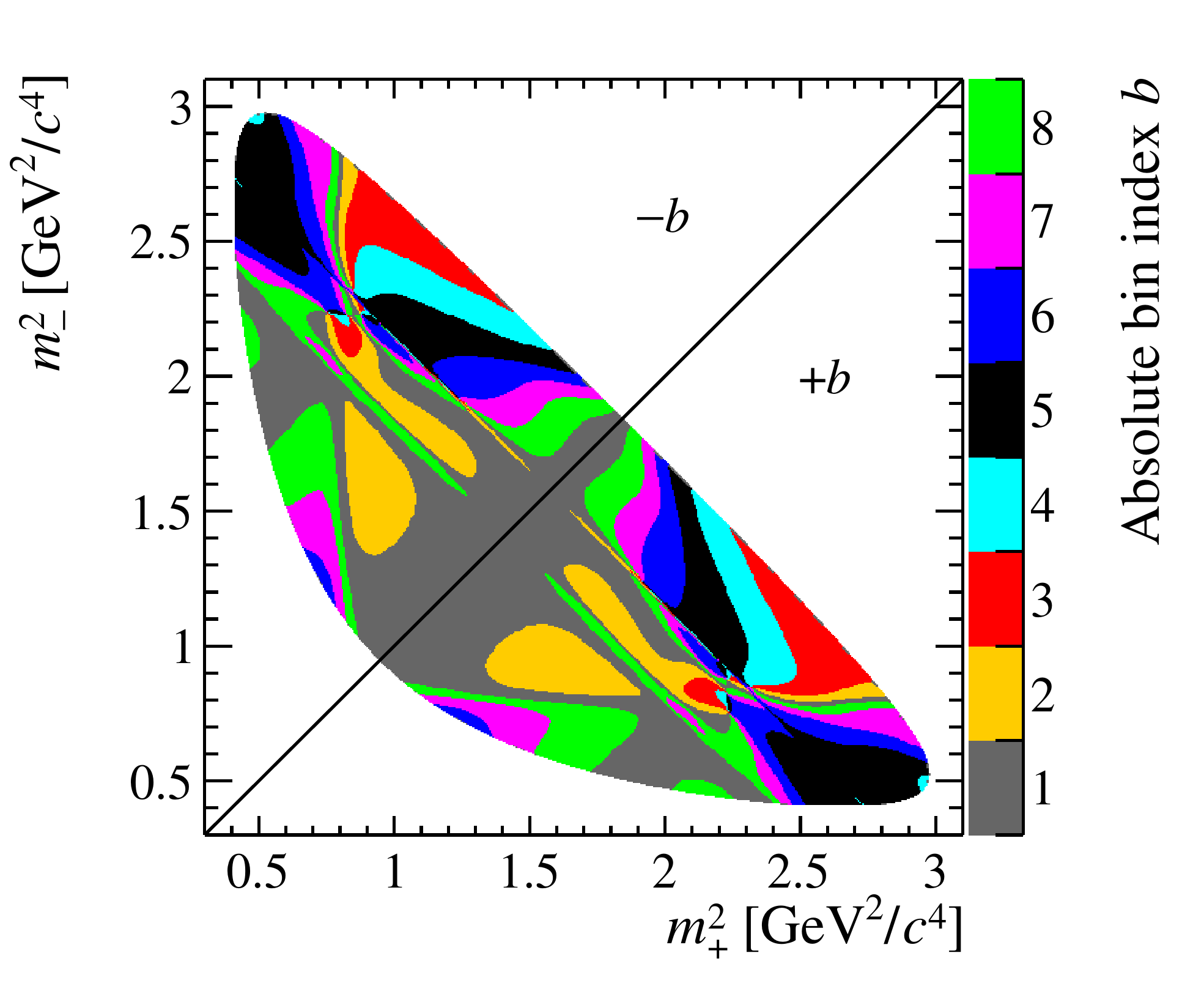}
\caption{Iso-$\Delta\delta$ binning of the \Dkspp Dalitz plot, based on the BaBar 2008 amplitude model~\cite{Libby:2010nu}. The bins are symmetric with respect to the $m_+^2=m_-^2$ bisector; positive indices refer to bins in the (lower) $m_+^2 > m_-^2$ region; negative indices refer to those in the (upper) $m_+^2 < m_-^2$ region. Colors indicate the absolute value of the bin index $b$.\label{fig:equalbinning}}
\end{figure}

The iso-$\Delta\delta$ scheme, shown in \cref{fig:equalbinning}, is available as a look-up table consisting of a grid of $(m_+^2, m_-^2)$ points spaced $0.0054\gevgevcccc$ apart in both $m_+^2$ and $m_-^2$. The corresponding values of $r_b$, $c_b$, and $s_b$ are reproduced in \cref{tab:rb-cs}, as measured by CLEO in $0.8\invfb$ of $e^+e^-$ collisions at a center-of-mass energy of $3.77\gev$. The $c_b$ and $s_b$ correlations are reported in \cref{tab:cs-corr}. 

\begin{table}[ht]
\centering
\caption{Values of $r_b$, $c_b$, and $s_b$ measured by CLEO for the BaBar 2008 iso-$\Delta\delta$ binning scheme~\cite{Libby:2010nu}. The first contribution to the uncertainty is statistical, the second systematic.\label{tab:rb-cs}}
\begin{tabular}{lrrr}
\toprule
$b$ & \multicolumn{1}{c}{$r_b$} & \multicolumn{1}{c}{$c_{b}$} & \multicolumn{1}{c}{$s_{b}$}\\
\midrule
1 & $0.488\pm0.028$ & $ 0.655\pm0.036\pm0.042$ & $-0.025\pm0.098\pm0.043$ \\
2 & $0.286\pm0.027$ & $ 0.511\pm0.068\pm0.063$ & $ 0.141\pm0.183\pm0.066$ \\
3 & $0.319\pm0.031$ & $ 0.024\pm0.140\pm0.080$ & $ 1.111\pm0.131\pm0.044$ \\
4 & $0.667\pm0.100$ & $-0.569\pm0.118\pm0.098$ & $ 0.328\pm0.202\pm0.072$ \\
5 & $0.632\pm0.052$ & $-0.903\pm0.045\pm0.042$ & $-0.181\pm0.131\pm0.026$ \\
6 & $0.220\pm0.036$ & $-0.616\pm0.103\pm0.072$ & $-0.520\pm0.196\pm0.059$ \\
7 & $0.125\pm0.016$ & $ 0.100\pm0.106\pm0.124$ & $-1.129\pm0.120\pm0.096$ \\
8 & $0.238\pm0.018$ & $ 0.422\pm0.069\pm0.075$ & $-0.350\pm0.151\pm0.045$ \\
\bottomrule
\end{tabular}
\end{table}

\begin{table}[ht]
\centering
\caption{Correlation coefficients (in \%) between the $c_{b}$ and $s_{b}$ parameters, separately for (top) statistical and (bottom) systematic effects, as measured by CLEO for the BaBar 2008 iso-$\Delta\delta$ binning scheme~\cite{Libby:2010nu}.\label{tab:cs-corr}}
\begin{tabular}{l@{\hspace{6pt}}r@{\hspace{6pt}}r@{\hspace{6pt}}r@{\hspace{6pt}}
r@{\hspace{6pt}}r@{\hspace{6pt}}r@{\hspace{6pt}}r@{\hspace{6pt}}r@{\hspace{6pt}}r@{\hspace{6pt}}
r@{\hspace{6pt}}r@{\hspace{6pt}}r@{\hspace{6pt}}r@{\hspace{6pt}}r@{\hspace{6pt}}r}
\toprule
 & $c_2$& $c_3$& $c_4$& $c_5$& $c_6$& $c_7$& $c_8$& $s_1$& $s_2$& $s_3$& $s_4$& $s_5$& $s_6$& $s_7$& $s_8$\\ \hline 
$c_1$& $-2$& $-3$& 5& 7& 3& 1& $-2$& 0& 0& $-2$& 0& 0& 0& $-1$& 0\\
$c_2$& & 0& 4& 10& 0& 0& 0& 0& $-2$& 0& 0& 0& 0& 0& 0\\
$c_3$& & & 0& 0& 0& 2& $-4$& 16& $-4$& 75& 7& $-10$& 0& 45& 4\\
$c_4$& & & & 1& 0& 0& 5& 0& $-1$& 0& 7& $-1$& 0& 0& 0\\
$c_5$& & & & & 0& 1& 2& 0& 3& 0& 0& 3& 0& 0& 0\\
$c_6$& & & & & & $-1$& $-1$& 1& 0& 0& 0& 0& 0& 0& 0\\
$c_7$& & & & & & & 0& 2& 3& 6& 0& 2& 0& 1& 2\\
$c_8$& & & & & & & & $-1$& 0& $-3$& 0& 0& 0& $-2$& 2\\
$s_1$& & & & & & & & & $-8$& 18& 11& $-18$& $-7$& 15& 10\\
$s_2$& & & & & & & & & & $-3$& 10& 31& $-6$& $-2$& 0\\
$s_3$& & & & & & & & & & & 11& $-9$& $-2$& 59& 6\\
$s_4$& & & & & & & & & & & & 0& $-4$& 13& 13\\
$s_5$& & & & & & & & & & & & & 6& $-10$& $-11$\\
$s_6$& & & & & & & & & & & & & & $-5$& $-6$\\
$s_7$& & & & & & & & & & & & & & & 3\\
\midrule
$c_1$& $\phantom{-}89$ & $\phantom{-}93$& $\phantom{-}74$& $\phantom{-}77$& $\phantom{-}85$& $\phantom{-}90$& $\phantom{-}90$& $\phantom{-}32$& $\phantom{-}24$& $\phantom{-}32$& $\phantom{-}30$& $\phantom{-}25$& $-11$& $\phantom{-}11$& $\phantom{-}29$\\
$c_2$& & 88& 70& 73& 83& 87& 90& 32& 25& 33& 33& 25& $-13$& 15& 28\\
$c_3$& & & 73& 77& 86& 91& 91& 34& 22& 37& 31& 23& $-9$& 13& 29\\
$c_4$& & & & 90& 80& 84& 79& $-11$& $-22$& 13& $-12$& 0& 24& $-31$& $-1$\\
$c_5$& & & & & 82& 83& 81& $-5$& $-14$& 16& $-6$& $-1$& 16& $-23$& 2\\
$c_6$& & & & & & 87& 87& 12& 7& 26& 15& 12& 4& $-2$& 17\\
$c_7$& & & & & & & 91& 17& 6& 24& 15& 15& 3& $-5$& 16\\
$c_8$& & & & & & & & 24& 15& 29& 24& 19& $-4$& 4& 20\\
$s_1$& & & & & & & & & 60& 37& 57& 29& $-43$& 58& 48\\
$s_2$& & & & & & & & & & 31& 55& 45& $-41$& 67& 51\\
$s_3$& & & & & & & & & & & 31& 23& $-9$& 35& 40\\
$s_4$& & & & & & & & & & & & 30& $-42$& 66& 49\\
$s_5$& & & & & & & & & & & & & $-20$& 27& 34\\
$s_6$& & & & & & & & & & & & & & $-56$& $-28$\\
$s_7$& & & & & & & & & & & & & & & 40\\
\bottomrule
\end{tabular}
\end{table}

\section{Sensitivity\label{sec:sensitivity}}
The bin-flip method is validated using simulated experiments. In \cref{sec:sensitivity:reach}, we discuss the tests of the basic assumptions and approximations of the method, study its properties, and offer an estimate of the best statistical precision possibly achievable. In \cref{sec:sensitivity:inputs}, we focus on the dependence of the method's sensitivity on external inputs. In \cref{sec:sensitivity:detectoreffects}, we discuss the impact of experimental effects, such as finite resolutions and nonuniform reconstruction efficiencies.

\subsection{Reach and comparison with other methods\label{sec:sensitivity:reach}}
\begin{figure}[t]
\centering
\includegraphics[width=0.6\textwidth]{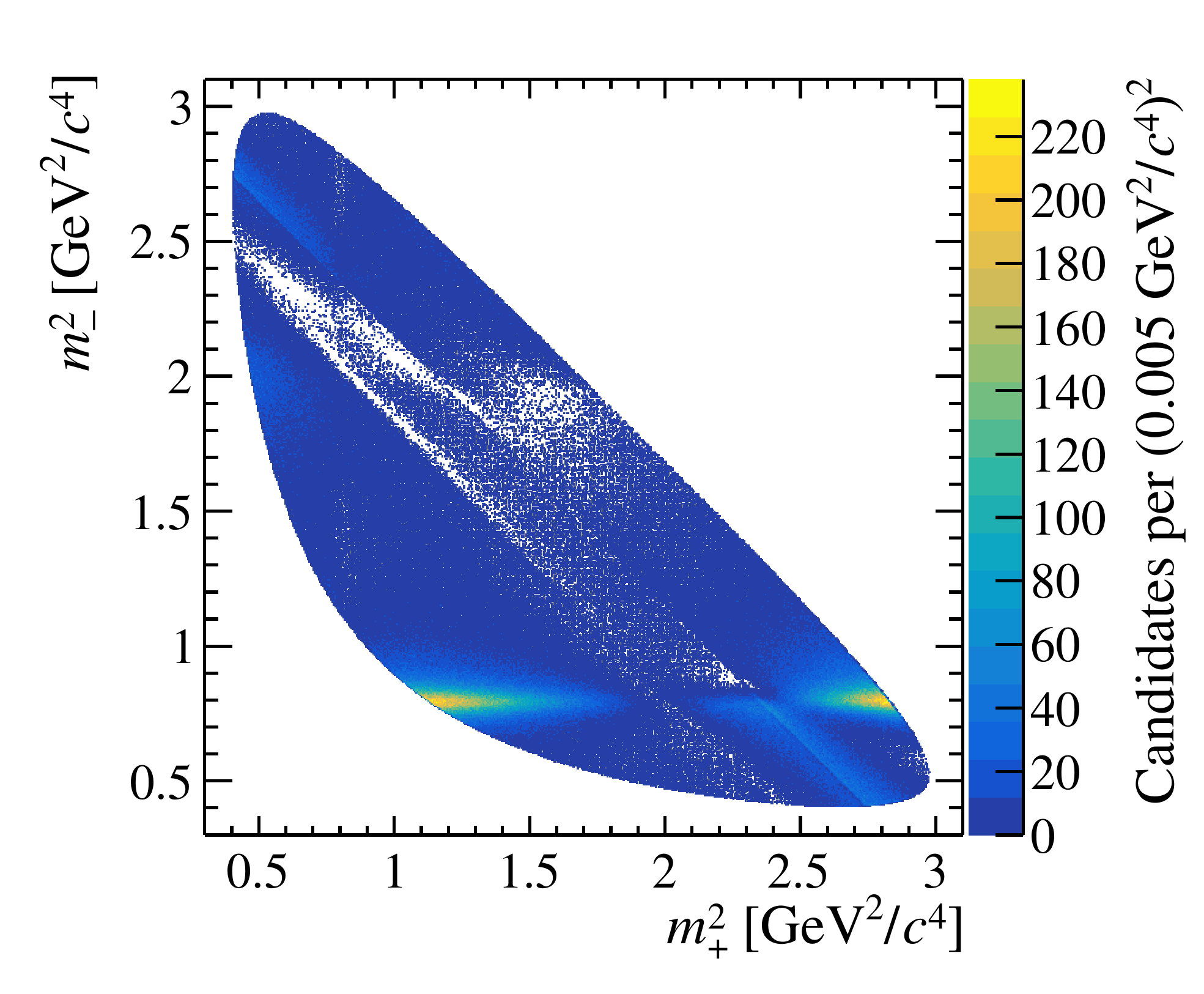}\\
\caption{\label{fig:example-toy} Dalitz-plot distribution for $10^6$ simulated \Dkspp decays in the no-mixing scenario.}
\end{figure}

The sensitivity of the bin-flip method to oscillation and \CP-violation parameters in charm mixing is determined using ensembles of simulated samples of \mbox{\Dkspp} decays, generated assuming five relevant configurations of the true values of such parameters:
\begin{enumerate}
\item[(i)] No mixing (NM), corresponding to $x=\xcp=y=\ycp=0$, $|q/p|=1$, and $\phi=0$ (or $\deltax=\deltay=0$);
\item[(ii)] \CP-conserving world-average mixing (WM), corresponding to $x=\xcp=0.4\%$, $y=\ycp=0.6\%$, $|q/p|=1$, and $\phi=0$ (or $\deltax=\deltay=0$);
\item[(iii)] \CP-conserving large mixing (LM), corresponding to $x=\xcp=y=\ycp=1\%$, $|q/p|=1$, and $\phi=0$ (or $\deltax=\deltay=0$);
\item[(iv)] \CP-violating world-average mixing (WCP), corresponding to $x=0.4\%$, $y=0.6\%$, $|q/p|=0.93$, and $\phi=-0.15$;
\item[(v)] World-average mixing with \CP violation in mixing only (MCP), corresponding to $x=0.4\%$, $y=0.6\%$, $|q/p|=0.93$, and $\phi=0$.
\end{enumerate}
The \Dkspp decays are generated by sampling the decay-time-dependent decay rate of \cref{eq:rate-D0,eq:rate-D0bar}~\cite{cfit}. The ``BaBar 2010'' model is used to describe the amplitudes at $t=0$ assuming \CP-conserving decay amplitudes (\cref{fig:example-toy})~\cite{delAmoSanchez:2010xz}. Minor differences between the amplitude model used in generation and the model used to define the iso-$\Delta\delta$ Dalitz-plot bins are irrelevant for testing the method. While the definition of bins requires an amplitude model, the method remains unbiased against mismodeling~\cite{Giri:2003ty,Bondar:2005ki,Bondar:2008hh,Bondar:2010qs,Thomas:2012qf}. In addition, both considered models achieve similar descriptions of the variations of the strong-interaction phases across phase-space, thus keeping the sensitivity of the method unaltered~\cite{Libby:2010nu}.

For each scenario we generate an ensemble of $10^3$ simulated experiments, each containing typically $10^6$ signal events, corresponding to the size of the full Belle sample and of that collected by LHCb during 2011--2012. No background or detector effects are simulated. Each sample is split into ten equally-populated decay-time bins and eight pairs of iso-$\Delta\delta$ Dalitz-plot bins. The average decay times and squared decay times are calculated in each bin using decays populating the lower Dalitz subspace ($m^2_-<m^2_+$, $b>0$) only, which is enriched in \Dz mesons that did not oscillate. The fit minimizes the least-squares function
\begin{equation}\label{eq:chi2}
\chi^2 = \sum_{b=1,j=1}^{b=8,j=10}\left[\frac{(N_{-bj}-N_{bj}R_{bj})^2}{(\sigma_{-bj})^2+(\sigma_{bj}R_{bj})^2}+\frac{(\Nbar_{-bj}-\Nbar_{bj}\Rbar_{bj})^2}{(\bar{\sigma}_{-bj})^2+(\bar{\sigma}_{bj}\Rbar_{bj})^2}\right].
\end{equation}
For each decay-time bin $j$ and pair of Dalitz-plot bin $\pm b$, the fit compares the decay yields $N_{\pm bj}$ ($\Nbar_{\pm bj}$) of charm mesons produced as \Dz (\Dzb) flavor states and observed in the chosen bin with the values expected from \cref{eq:bin-flip-ratio-D0,eq:bin-flip-ratio-D0bar} by weighting their squared difference with the variance, which is a function of the yield's uncertainties $\sigma_{\pm bj}$ ($\bar{\sigma}_{\pm bj}$). Signal yields follow Poisson distributions to a good approximation. Hence, we approximate the uncertainties of the yields as the square roots of the numbers of decays.

\begin{figure}[t]
\centering
\includegraphics[width=\textwidth]{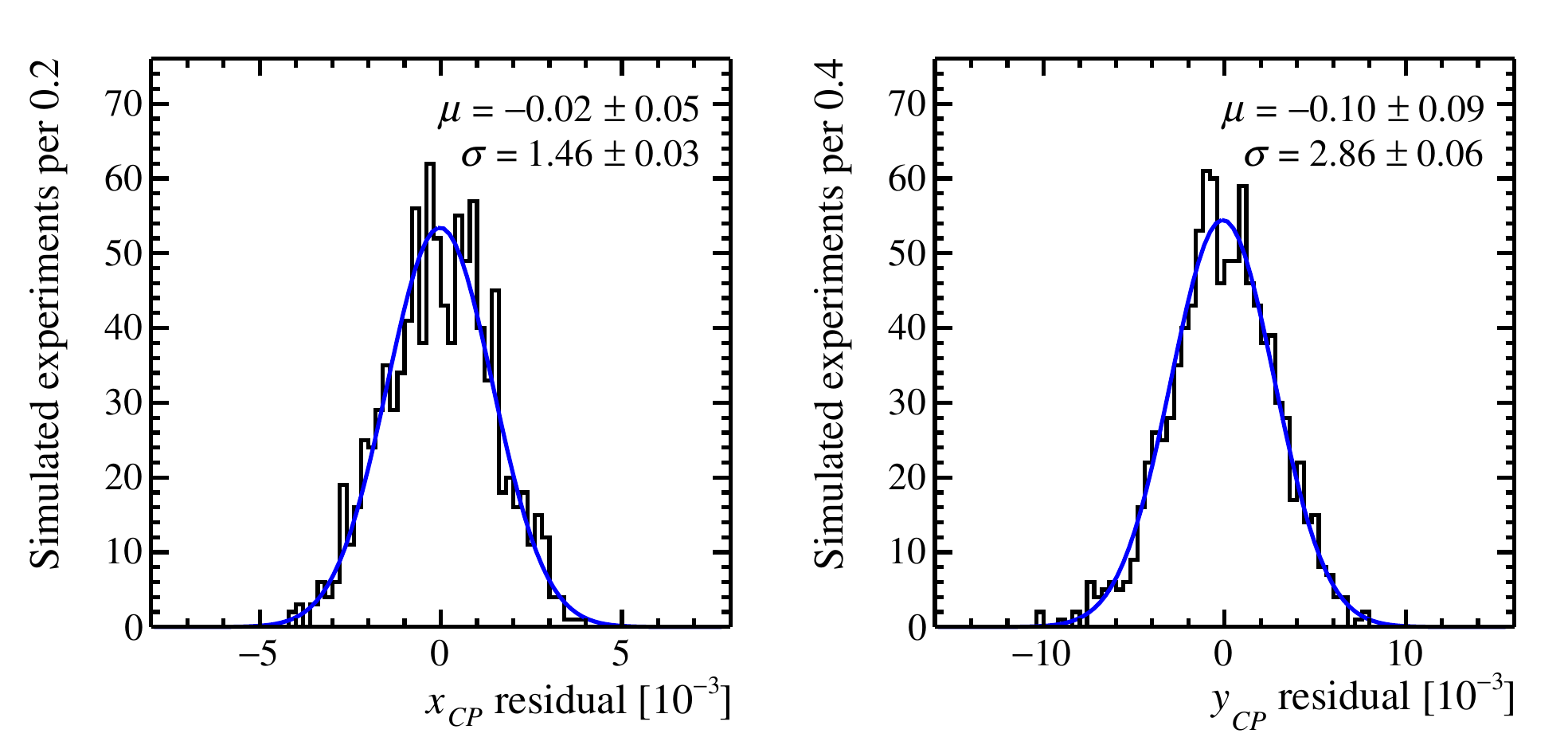}\\
\includegraphics[width=\textwidth]{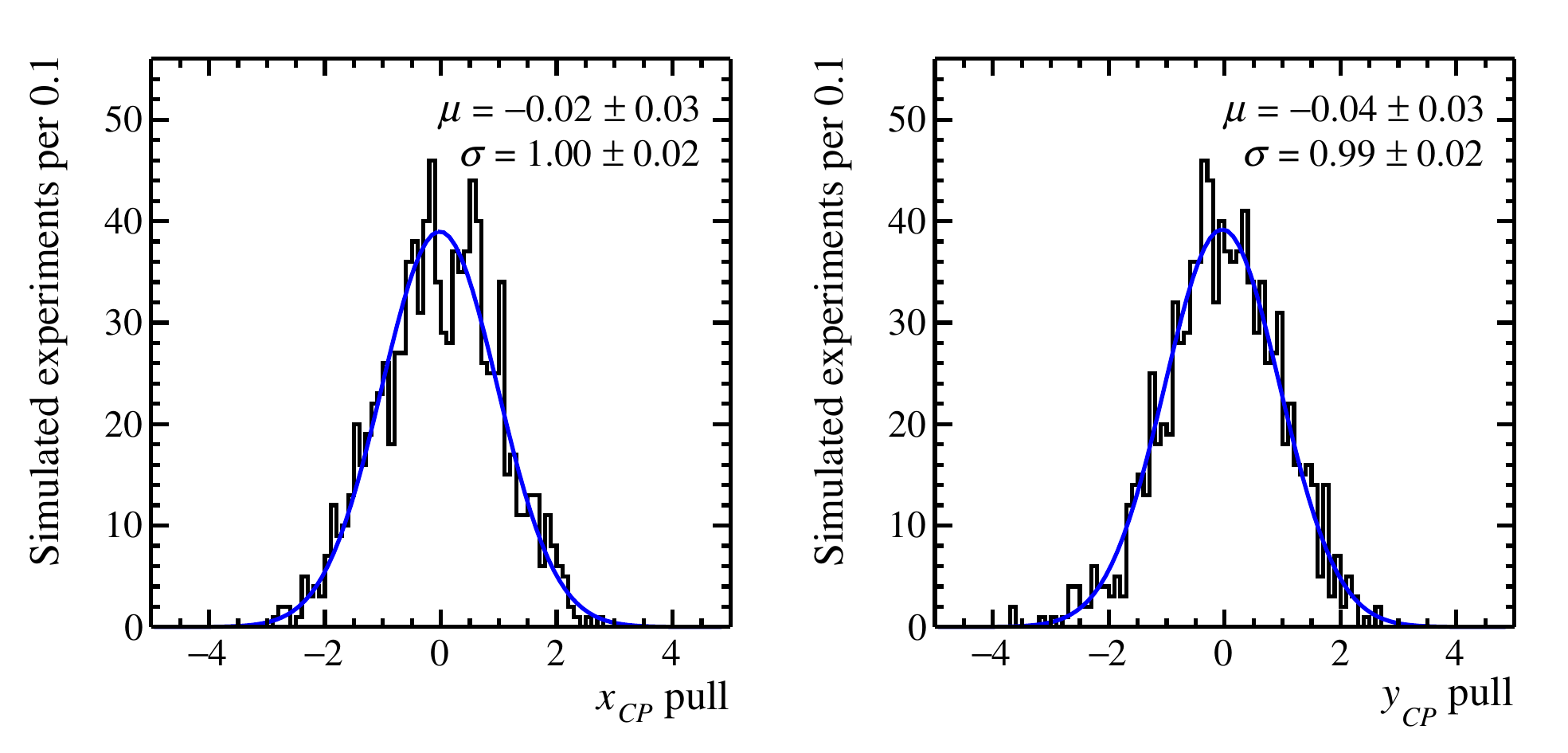}\\
\caption{Distributions of (top) fit residuals and (bottom) pulls on (left) \xcp and (right) \ycp from $10^6$ \Dkspp decays in the WM scenario, and fit assuming \CP conservation. Only the mixing parameters are determined by the fits.\label{fig:gen-toy-WM}}
\end{figure}

\cref{fig:gen-toy-WM} shows, as an example, the distribution of the difference between estimated and generated values (fit residual) and the distribution of the same difference divided by the estimated standard deviation (fit pull) obtained in fits assuming \CP conservation in the WM scenario, with nuisance parameters $r_b$, $c_b$, and $s_b$ fixed to their input values (\cref{tab:input-toys}). The estimated uncertainties on $\xcp=x$ and $\ycp=y$ are $0.15\%$ and $0.29\%$, respectively. The difference in sensitivity to $x$ and $y$ is due to the partial cancellation, in the yield ratio, of mixing terms proportional to amplitudes with defined \CP eigenvalues (such as the \CP-odd $\Dz\to\KS\rho(770)^0$ amplitude), which carry sensitivity to $y$. A coarse estimate of the expected reduction in sensitivity is obtained by further expanding the ratio of \cref{eq:bin-flip-ratio-D0} for $\langle t\rangle_j\sqrt{r_b}\re{X_b z}\ll1$ while retaining only terms linear in decay time, and under the simplifying assumption of \CP conservation,
\begin{align}\label{eq:approximate-bin-flip-ratio}
R_{bj} &\approx \frac{r_b+\langle t\rangle_j\sqrt{r_b}\re{X_b^\star z}}{1+\langle t\rangle_j\sqrt{r_b}\re{X_b z}}\nonumber\\
&\approx r_b+\langle t\rangle_j\sqrt{r_b}\bigl[\re{X_b^\star z} - r_b\re{X_b z}\bigr]\nonumber\\
&= r_b-\langle t\rangle_j\sqrt{r_b}\bigl[(1-r_b)c_b\ y - (1+r_b)s_b\ x\bigr].
\end{align}
The coefficient multiplying $y$ is typically half of that multiplying $x$ (\cref{tab:input-toys}) suggesting halved uncertainties on \xcp and \deltax with respect to those on \ycp and \deltay, respectively. In the limit of a Dalitz-plot bin saturated by \CP-eigenstate amplitudes, where $r_b \approx 1$ and $s_b \approx 0$, sensitivity to the mixing parameters vanishes.

\begin{table}[t]
\centering
\caption{\label{tab:input-toys} Values of $r_b$, $c_b$, and $s_b$ resulting from the BaBar 2010 amplitude model~\cite{delAmoSanchez:2010xz} used to generate the simulated experiments, with the BaBar 2008 iso-$\Delta\delta$ binning of the \Dkspp Dalitz plot defined by CLEO~\cite{Libby:2010nu}.}
\begin{tabular}{lrrr}
\toprule
$b$ & \multicolumn{1}{c}{$r_b$} & \multicolumn{1}{c}{$c_b$} & \multicolumn{1}{c}{$s_b$} \\
\midrule
  1	& $0.479784$ & $ 0.670828$ & $-0.032140$\\
  2	& $0.221027$ & $ 0.635411$ & $ 0.395893$\\
  3	& $0.276147$ & $ 0.087385$ & $ 0.850636$\\
  4	& $0.678943$ & $-0.490907$ & $ 0.783871$\\
  5	& $0.588435$ & $-0.946404$ & $ 0.113501$\\
  6	& $0.239850$ & $-0.681781$ & $-0.453785$\\
  7	& $0.107627$ & $-0.131118$ & $-0.813580$\\
  8	& $0.208639$ & $ 0.381420$ & $-0.482809$\\
\bottomrule
\end{tabular}
\end{table}

\begin{figure}[t]
\centering
\includegraphics[width=0.5\textwidth]{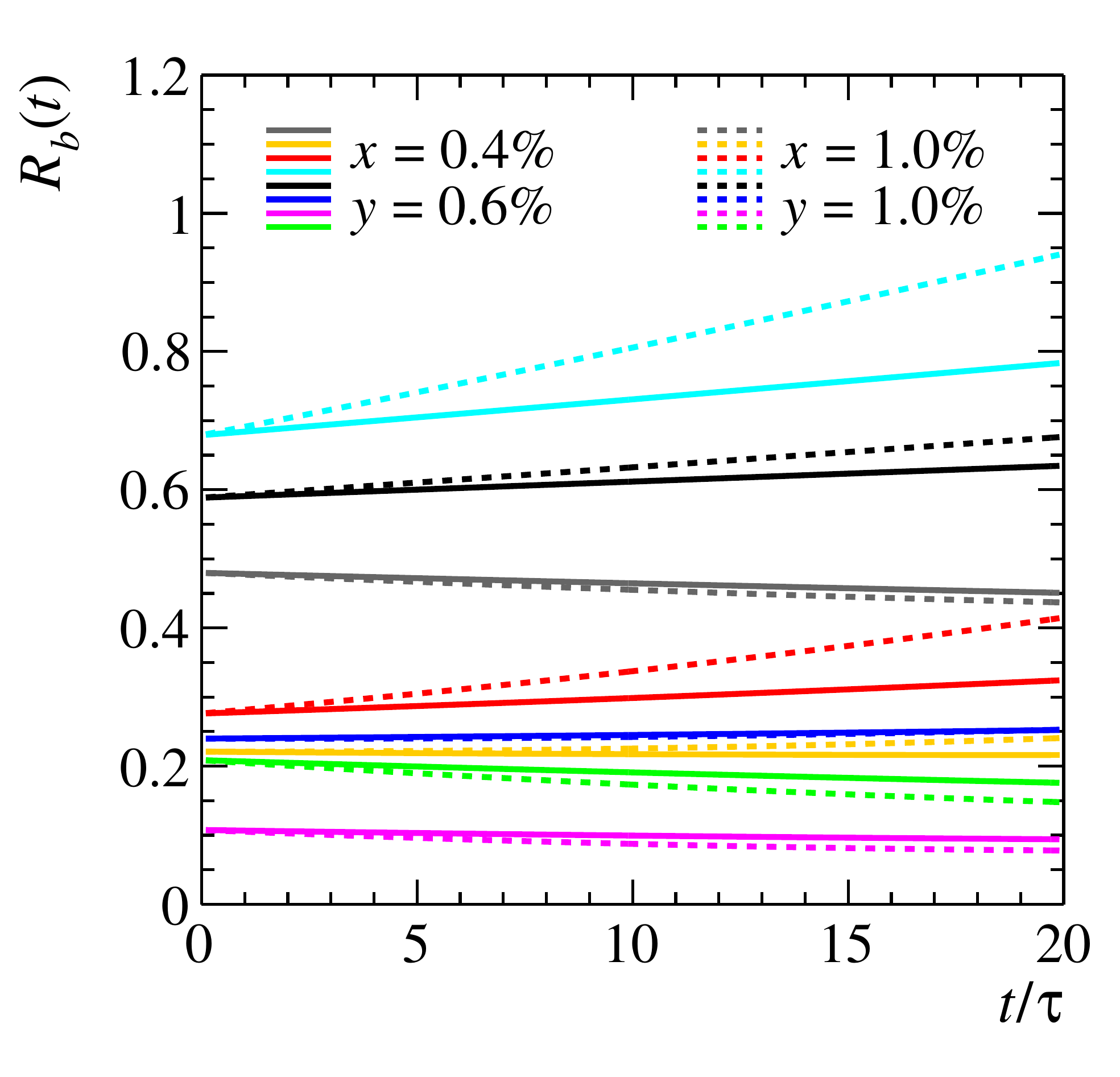}\\
\caption{Bin-flip ratio in each Dalitz-plot bin as a function of decay time, in the limit of \CP symmetry and for two sets of mixing parameters. Ratios are computed from \cref{eq:bin-flip-ratio-D0} with hadronic parameters from \cref{tab:input-toys}.  Colors identify the various Dalitz-plot bins following \cref{fig:equalbinning}.\label{fig:ratios-vs-t}}
\end{figure}

\Cref{eq:approximate-bin-flip-ratio} allows an illustration of the bin-flip method through an analogy with the wrong-sign-to-right-sign analysis of $\Dz\to K^{\mp}\pi^{\pm}$ decays~\cite{PhysRevLett.60.1239,pdg}. For the $\Dz\to K^{\mp}\pi^{\pm}$ analysis, a similar ratio is obtained, but with parameters that correspond to a single amplitude ratio rather than their averages over a Dalitz-plot bin.  Thus, $c_b$ and $s_b$ are replaced by $\cos \delta$ and $\sin \delta$, respectively, while the replacement for $r_b$ is conventionally indicated as $R_D$. Factors $(1 \pm R_D)$ are neglected since $R_D \ll 1$. The sign of the term linear in $\langle t\rangle$ is also flipped, due to a difference in the conventions to define $\delta$ that amounts to a shift of $\pi$. The mixing effect in the bin-flip method can therefore be visualized as slopes in the decay-time ($j$) dependences of $R_{bj}$ that are correlated between Dalitz-plot bins ($b$), as shown in \cref{fig:ratios-vs-t}. For bins where $r_b$ approaches 1, $s_b$ is large, and $c_b$ is small (\eg, bin~4 in \cref{tab:input-toys}), the effect is mainly due to $x$; for bins where $r_b$ and $s_b$ are small, but $c_b$ is large (\eg, bin~5), the effect is mainly due to $y$. Hence, the observed slopes and the known values of $c_b$ and $s_b$ allow for determining both $x$ and $y$, unlike in $\Dz\to K^{\mp}\pi^{\pm}$ decays, where only a single combination of $x$ and $y$ is accessible (since $\sin \delta$ is close to zero, the $\Dz\to K^{\mp}\pi^{\pm}$ analysis is primarily sensitive to $y$). Contributions from potential \CP-violation effects are inferred by comparing the slopes of the ratios for mesons produced as \Dz or \Dzb separately. 

\begin{table}[t]
\centering
\caption{Expected statistical uncertainties from $10^6$ simulated \Dkspp decays generated in the WM scenario, and analyzed with different approaches assuming \CP conservation and allowing only the mixing parameters to float in the fit.\label{tab:sensitivities-comparison}}
\begin{tabular}{lcc}
\toprule
Analysis method & $\sigma(x)$ [\%] & $\sigma(y)$ [\%]\\
\midrule
Model-dependent            & $0.11$ & $0.10$ \\
Standard model-independent & $0.20$ & $0.18$ \\
Bin-flip model-independent & $0.15$ & $0.29$ \\
\bottomrule
\end{tabular}
\end{table}

In addition, we use the simulated samples to compare the performance of the bin-flip method to those of existing approaches. The customary model-dependent analysis implies a joint maximum-likelihood fit to the unbinned decay-time and Dalitz-plot distributions, based on the same amplitude model used in generation. The established   model-independent analysis implies a joint maximum-likelihood fit to the unbinned decay-time distributions of decays in the 16 Dalitz-plot bins. While evaluating the performance of both standard methods, we keep all parameters fixed except $x$ and $y$. \Cref{tab:sensitivities-comparison} reports the results. Predictably, when the underlying amplitude model is exactly known, the model-dependent analysis offers the best sensitivity to both $x$ and $y$. The bin-flip method provides better sensitivity to $x$ than the known  model-independent method at the price of reduced sensitivity to $y$. This is expected because (i) the decay-time binning affects only marginally the statistical precision with $\mathcal{O}(10^6)$ signal yields or larger and (ii) the coefficients of the terms associated with sensitivity to $x$ and $y$ are enhanced or suppressed by $(1+r_b)$ and $(1-r_b)$, respectively, in the bin-flip method compared to the standard model-independent approach, as shown in \cref{eq:approximate-bin-flip-ratio}. By averaging over the Dalitz-plot bins, the coefficient multiplying $x$ ($y$) in the bin-flip method becomes approximately $35\%$ larger (smaller) than that from the original model-independent method, consistent with the sensitivities of \cref{tab:sensitivities-comparison}.

\begin{figure}[t]
\centering
\includegraphics[width=\textwidth]{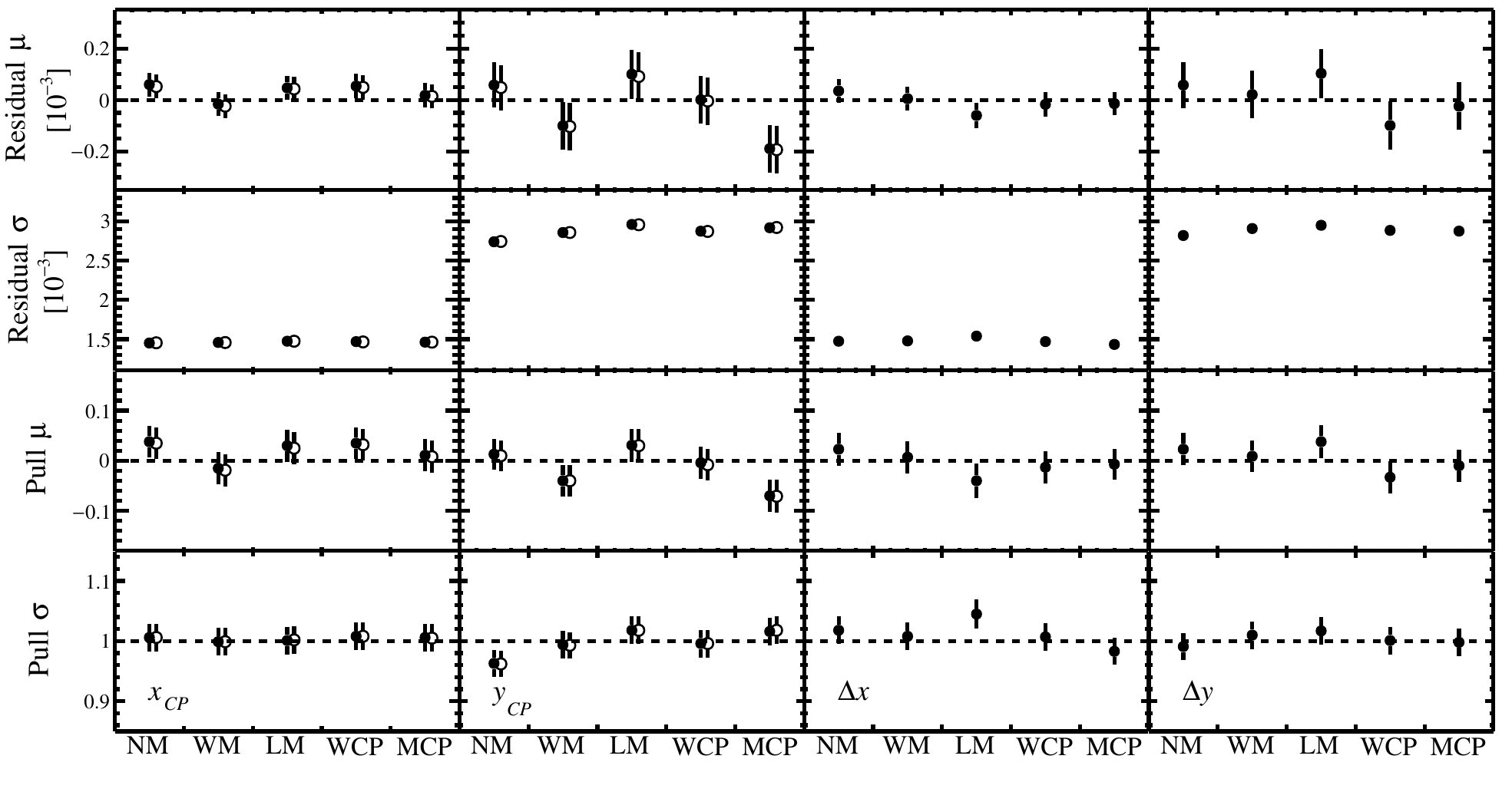}
\caption{Summary of results from simulated experiments of $10^6$ \Dkspp decays generated in the five scenarios of \CP-conserving or \CP-violating mixing, and fit (open points) assuming \CP conservation or (closed points) allowing for indirect \CP violation. Only the mixing and \CP-violation parameters float in the fits.\label{fig:gen-toy-summary}}
\end{figure}

\Cref{fig:gen-toy-summary} summarizes the results obtained in the various scenarios and for either assumption on indirect \CP violation. The estimates obtained with the bin-flip method are unbiased and show proper statistical uncertainties for all the relevant parameters, regardless of their true values. In addition, the precision on the \CP-averaged mixing parameters does not depend on whether the \CP-violation parameters are fixed or determined by the fit, which is expected since \zcp and \deltaz are additive, orthogonal parameters. The customary multiplicative parametrization in terms of $z^\pm=z(q/p)^{\pm1}$ yields larger correlations between mixing and indirect \CP-violation parameters, which bias the estimators and induce non-Gaussian uncertainties (\cref{app:cpvparametrization}). Indeed, the uncertainties on $|q/p|$ and $\phi$ obtained with the existing model-dependent method depend strongly on the estimated values of mixing parameters $x$ and $y$~\cite{Peng:2014oda}, which is undesirable, especially in combinations of results. The uncertainties on the parameters set out in \cref{eq:xcp-def,eq:dx-def,eq:ycp-def,eq:dy-def} do not depend on the central values of any of the other parameters, thus showing better statistical properties.

\begin{figure}[t]
\centering
\includegraphics[width=0.5\textwidth]{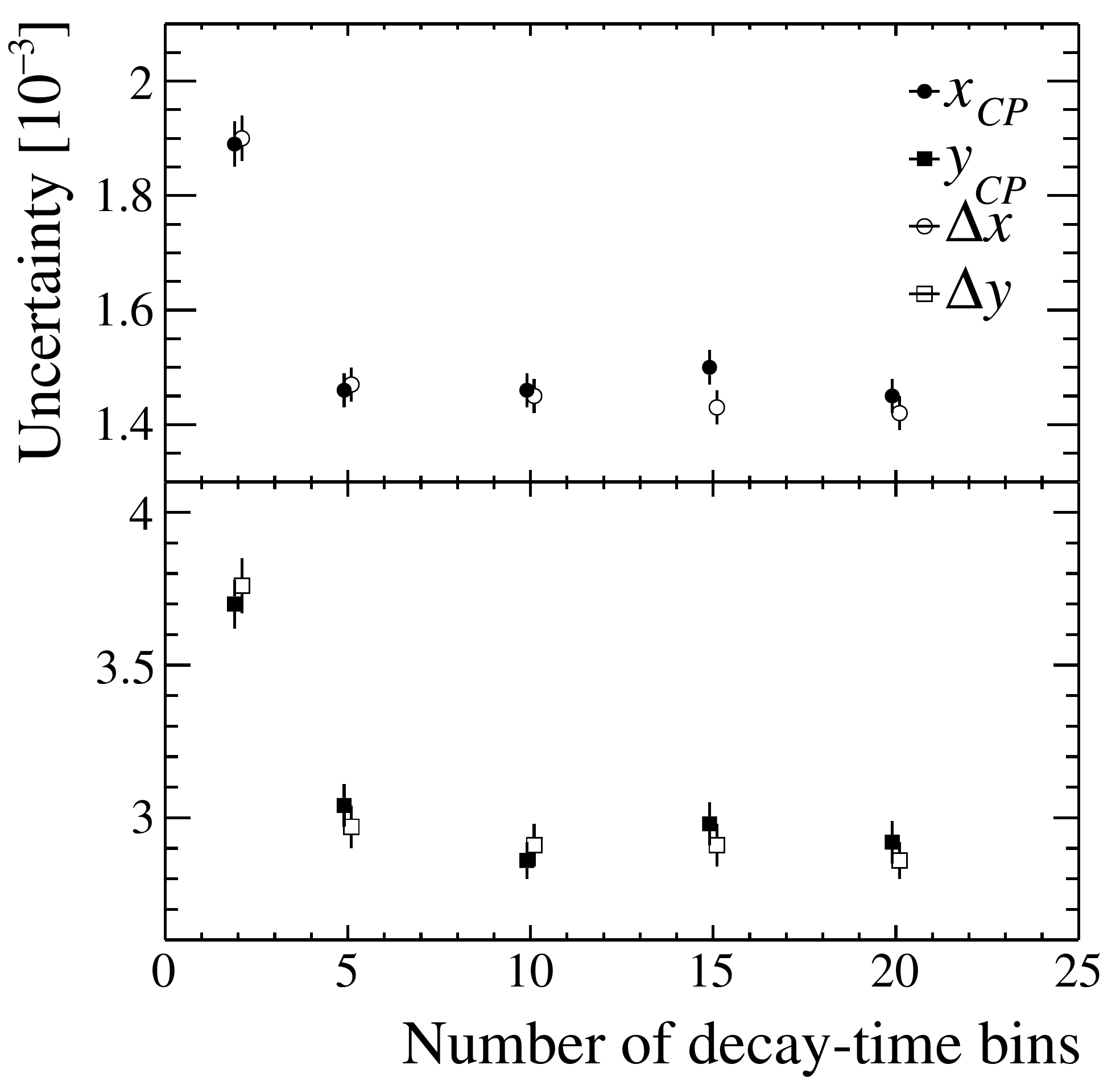}\\
\includegraphics[width=0.5\textwidth]{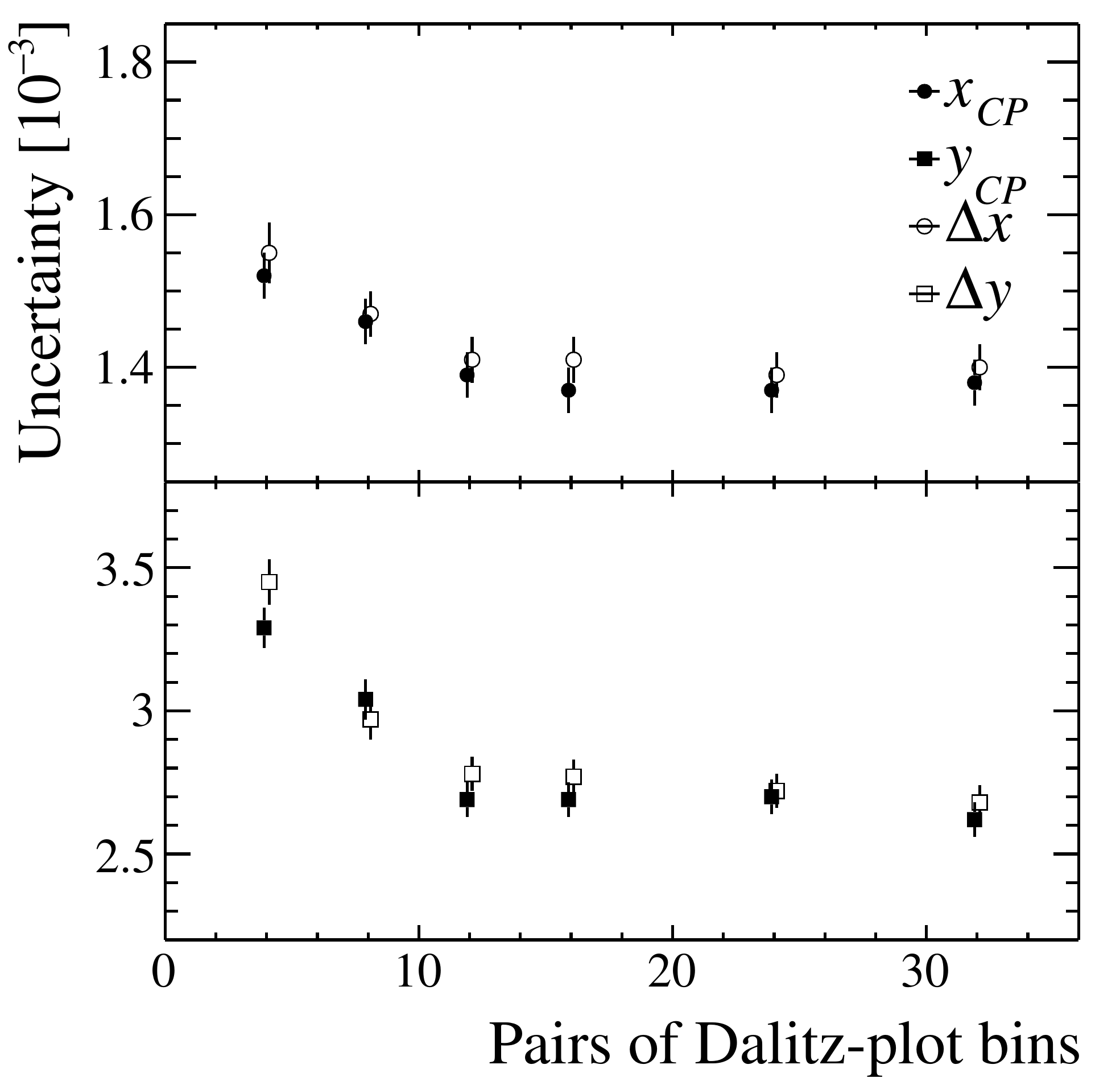}\\
\caption{Uncertainties on the mixing and \CP-violation parameters as functions of the number of (top) decay-time and (bottom) pairs of Dalitz-plot bins, as obtained from fitting simulated samples of $10^6$ \Dkspp decays each generated in the WM scenario. Only mixing and indirect \CP-violation parameters float in the fits.\label{fig:sigma-vs-nbins}}
\end{figure}

We study the dependence of our findings on  bin multiplicity by repeating the study with various choices for the number of decay-time bins and of pairs of Dalitz-plot bins. In all tests we consider equipopulated decay-time bins and iso-$\Delta\delta$ Dalitz-plot bins. \Cref{fig:sigma-vs-nbins} shows no significant dependence on the number of decay-time (pairs of Dalitz-plot) bins if that exceeds approximately five (ten). Since all schemes investigated experimentally thus far involve eight pairs of \Dkspp Dalitz-plot bins~\cite{Libby:2010nu}, alternative schemes that are either optimized for the bin-flip method or simply use more bins could result in greater sensitivity.

\subsection{Dependence on external inputs\label{sec:sensitivity:inputs}}
The sensitivity studies of \cref{sec:sensitivity:reach} assume the ideal case in which the hadronic parameters $r_b$, $c_b$, and $s_b$ are known exactly. A more realistic scenario, however, requires these parameters to be either determined directly from the data, or constrained by external measurements. Since the samples of $e^+e^-$ collisions at the $\psi(3770)$ resonance used to determine the strong-interaction phase parameters are, and will be, smaller than those used in upcoming measurements of charm-mixing parameters in \mbox{\Dkspp} decays, independent higher-precision measurements to constrain $r_b$ will be lacking. Moreover, biases due to efficiency variations across the Dalitz plot are likely to become non-negligible at the precision expected on $r_b$. It is therefore convenient to leave $r_b$ as a free parameter in the fit.  More precisely, the free parameter is $\tilde{r}_b = \tilde{F}_{-b}/\tilde{F}_b$, following \cref{eq:Ftilde}, and is treated as an effective, nuisance parameter that has no straightforward interpretation in terms of the underlying amplitude models. Conversely, since the mixing and $(c_b,s_b)$ parameters cannot simultaneously be determined precisely without external inputs, an appropriate binning scheme and care over the efficiency modeling is required to keep the biases on $c_b$ and $s_b$ minimal (see \cref{sec:sensitivity:detectoreffects}). Therefore, the optimal fit configuration for a realistic analysis corresponds to keeping $r_b$ free to vary and $(c_b,s_b)$ constrained.

\begin{table}[b]
\centering
\caption{Expected statistical sensitivities from $10^6$ simulated \Dkspp decays generated in the WM scenario, and fit under various assumptions. The fit configuration that best approximates the conditions in a realistic analysis corresponds to allowing $r_b$ to float free and keeping $c_b$ and $s_b$ constrained (last row in each subpanel).\label{tab:gen-toy-uncertainties}}
\begin{tabular}{lcccc}
\toprule
Fit configuration & $\sigma(\xcp)$ [\%]& $\sigma(\ycp)$ [\%] & $\sigma(\deltax)$ [\%]& $\sigma(\deltay)$ [\%]\\
\midrule
No \CP violation \\
$\quad r_b$, $(c_b,s_b)$ fixed             & $0.15$ & $0.29$ & -- & -- \\
$\quad r_b$ free, $(c_b,s_b)$ fixed        & $0.21$ & $0.41$ & -- & -- \\
$\quad r_b$ fixed, $(c_b,s_b)$ constrained & $0.16$ & $0.30$ & -- & -- \\
$\quad r_b$ free, $(c_b,s_b)$ constrained  & $0.22$ & $0.43$ & -- & -- \\
\midrule
Indirect \CP violation allowed \\
$\quad r_b$, $(c_b,s_b)$ fixed             & $0.15$ & $0.29$ & $0.15$ & $0.29$ \\
$\quad r_b$ free, $(c_b,s_b)$ fixed        & $0.21$ & $0.41$ & $0.15$ & $0.29$ \\
$\quad r_b$ fixed, $(c_b,s_b)$ constrained & $0.16$ & $0.30$ & $0.16$ & $0.31$ \\
$\quad r_b$ free, $(c_b,s_b)$ constrained  & $0.22$ & $0.43$ & $0.16$ & $0.31$ \\
\bottomrule
\end{tabular}
\end{table}

\Cref{tab:gen-toy-uncertainties} shows the sensitivity of the uncertainties to the choice of fit configuration (unconstrained or constrained) for the nuisance parameters $r_b$, $c_b$, and $s_b$. The constraints on $(c_b,s_b)$ are implemented by adding to \cref{eq:chi2} the penalty term 
\begin{equation}
\chi^2_{X} = \sum_{a,b}\left[X^{\rm{gen}}_a-X_a\right](V_{\rm{CLEO}}^{-1})_{ab}\left[X^{\rm{gen}}_b-X_b\right],
\end{equation}
where $X_b^{\rm{gen}}$ are the generator-level values of \cref{tab:input-toys} and the covariance matrix $V_{\rm{CLEO}}$ is the sum of the statistical and systematic covariance matrices from the CLEO measurement of $(c_b,s_b)$ derived from the values reproduced in \cref{tab:rb-cs,tab:cs-corr}~\cite{Libby:2010nu}. With $10^6$ \mbox{\Dkspp} decays, the impact of the current precision of measurements of $(c_b,s_b)$ is marginal. If $r_b$ is unconstrained in the fit, a more significant impact on $\sigma(\xcp)$ and $\sigma(\ycp)$ is expected whereas $\sigma(\deltax)$ and $\sigma(\deltay)$ are unaffected. 

To assess the impact of the limited precision of external constraints on future larger samples of \Dkspp decays, such as those expected at the LHCb and Belle~II experiments, the sensitivity is evaluated as a function of sample size. LHCb is expected to collect about $5\times10^7$ decays by the end of 2018 and at least an order of magnitude more by 2030, after detector upgrades~\cite{Bediaga:2018lhg}. Belle~II is expected to collect about $1\times10^6$ decays per 1\,ab$^{-1}$ of integrated luminosity, for a total of about $5\times10^7$ decays by the end of 2025~\cite{Kou:2018nap}. \Cref{tab:gen-toy-uncertainties-vs-n} shows uncertainties on the oscillation and \CP-violation parameters in charm mixing resulting from fits with unconstrained $r_b$ parameters and $(c_b,s_b)$ either constrained or fixed. The precision of currently available measurements of $(c_b,s_b)$ from CLEO will start impacting the precision on \xcp and \ycp with $10^7$ decays, but has negligible impact on the determination of the \CP-violation parameters \deltax and \deltay. However, more precise inputs are expected owing to $\mathcal{O}(10)$ times larger data sets of $e^+e^-$ collisions at center-of-mass energy of $3.77\gev$ that are being collected with the BESIII detector at the Beijing Electron-Positron Collider. It is therefore plausible to expect that the uncertainty due to external inputs will reduce, mirroring the reduction in statistical uncertainty and thus not limiting the precision of the proposed method.

\begin{table}[t]
\centering
\caption{Expected statistical uncertainties as functions of signal yields from fits to simulated \Dkspp decays generated in the WM scenario, allowing $r_b$ to float freely and keeping $(c_b,s_b)$ constrained (fixed). The constraints are based on the uncertainties of the CLEO results~\cite{Libby:2010nu}.\label{tab:gen-toy-uncertainties-vs-n}}
\begin{tabular}{lcccc}
\toprule
Signal yield & $\sigma(\xcp)$ [\%]& $\sigma(\ycp)$ [\%] & $\sigma(\deltax)$ [\%]& $\sigma(\deltay)$ [\%]\\
\midrule
$1\times10^6$ & $0.22\0\,(0.21\0)$ & $0.43\0\,(0.41\0)$ & $ 0.16\0\,(0.15\0)$ & $0.31\0\,(0.29\0)$ \\
$5\times10^6$ & $0.10\0\,(0.093)$ & $0.24\0\,(0.19\0)$ & $0.068\,(0.065)$ & $0.16\0\,(0.13\0)$ \\
$1\times10^7$ & $0.085\,(0.066)$ & $0.16\0\,(0.13\0)$ & $0.048\,(0.046)$ & $0.095\,(0.091)$ \\
$5\times10^7$ & $0.047\,(0.030)$ & $0.120\,(0.059)$ & $0.021\,(0.021)$ & $0.041\,(0.041)$ \\
$1\times10^8$ & $0.043\,(0.021)$ & $0.091\,(0.042)$ & $0.015\,(0.015)$ & $0.028\,(0.028)$ \\
$5\times10^8$ & $0.034\,(0.009)$ & $0.091\,(0.018)$ & $0.006\,(0.006)$ & $0.013\,(0.013)$ \\
\bottomrule
\end{tabular}
\end{table}

The above analysis is carried out in the limit of \CP-symmetric \D decay amplitudes. As larger data sets will become available, this assumption might need to be revisited, possibly resulting in an extension of the method toward including direct \CP asymmetries as has been considered for the GGSZ method~\cite{Bondar:2013jxa}. We expect that doing so will enrich the physics reach of the method without significantly affecting the sensitivity to oscillation and indirect \CP violation.

\subsection{Effects of finite resolutions and nonuniform efficiencies\label{sec:sensitivity:detectoreffects}}
For the bin-flip method to be applicable to experimental data, effects such as backgrounds, flavor tagging, finite resolutions, and nonuniform efficiency variations across decay time and Dalitz plane need in principle to be accounted for. Backgrounds and flavor tagging are not a significant limitation. Using the $\Dstar(2010)^+ \to \Dz\pip$ decay chain provides both very effective background rejection and a highly efficient and pure identification of the initial \D meson flavor. Reconstruction effects can also be accounted for, by weighting the candidates by the inverse of the efficiency at a given point in phase space and decay time, for example. However, the determination of the detector resolution and efficiency variations often relies on an accurate simulation of the detector response, which may introduce further unwanted sources of systematic uncertainties and complexity in the analysis procedures.

The bin-flip method is constructed so as to be insensitive to such effects. To validate this notion, we incorporate in the simulated samples realistic resolution and efficiency effects based on publicly available information from the LHCb and Belle~II experiments, which are the environments where this method is most likely to be considered. In both cases we consider experimental effects typical of $\Dz \to \KS (\to \pip\pim)\pip\pim$ signal decays reconstructed from $\Dstar(2010)^+\to\Dz\pip$ decays. At LHCb, significant samples of \Dz mesons are also obtained from semileptonic $B$-meson decays, with online selection-requirements that induce less distortion of the kinematic and decay-time distributions. Such samples can therefore provide results on charm-mixing parameters complementary to those based on \Dz mesons produced at the proton-proton primary interaction~\cite{LHCb-PAPER-2016-033}. These are not considered in this work.

\begin{figure}[b]
\centering
\includegraphics[width=0.5\textwidth]{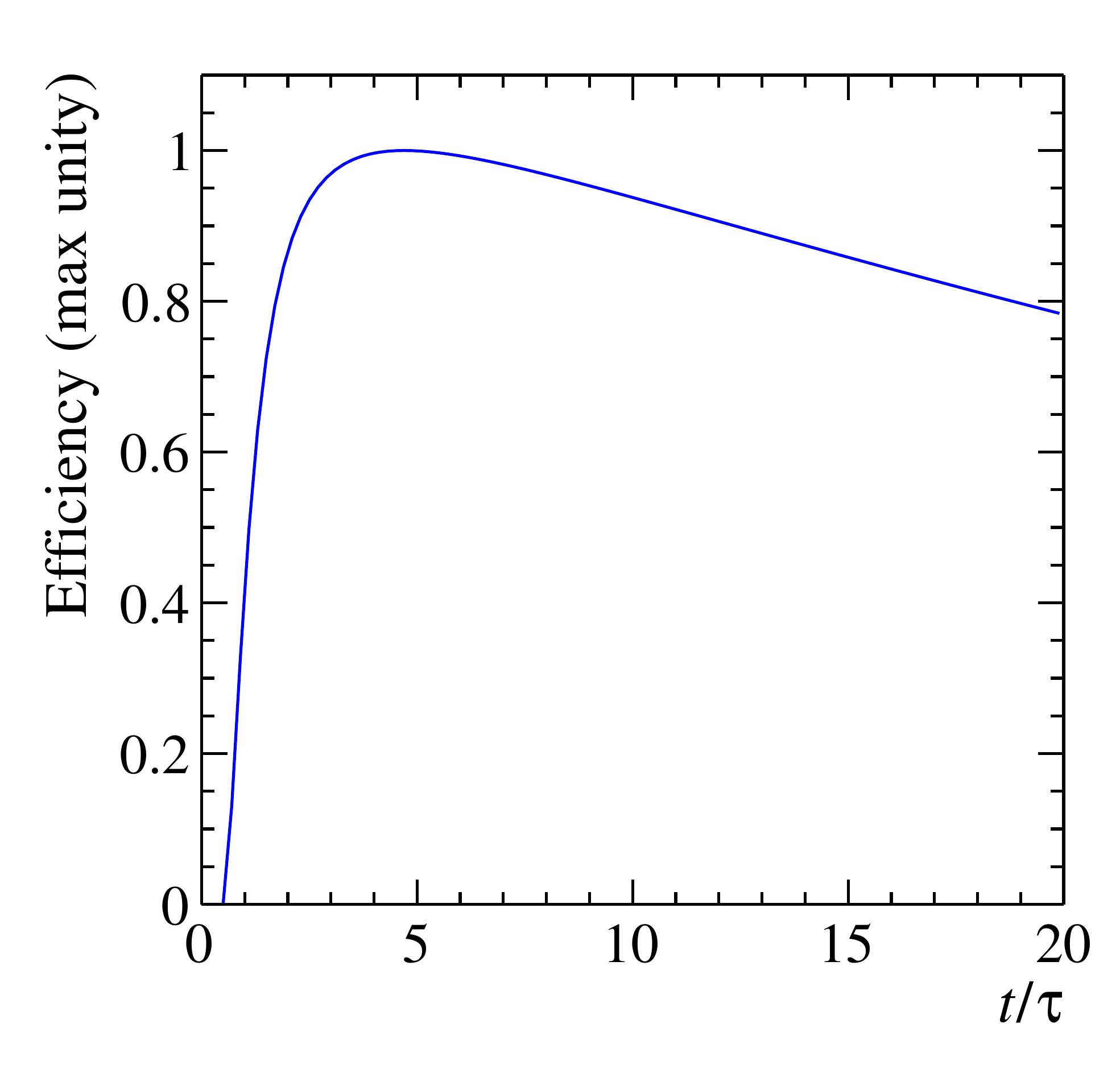}
\caption{Efficiency (normalized to unity at its maximum) as a function of decay time assumed for \mbox{\Dkspp} decays reconstructed from the $\Dstar(2010)^+\to\Dz\pip$ decay chain with the LHCb detector.\label{fig:prompt_dt_acc}}
\end{figure}

\begin{figure}[ht]
\centering
\includegraphics[width=0.6\textwidth]{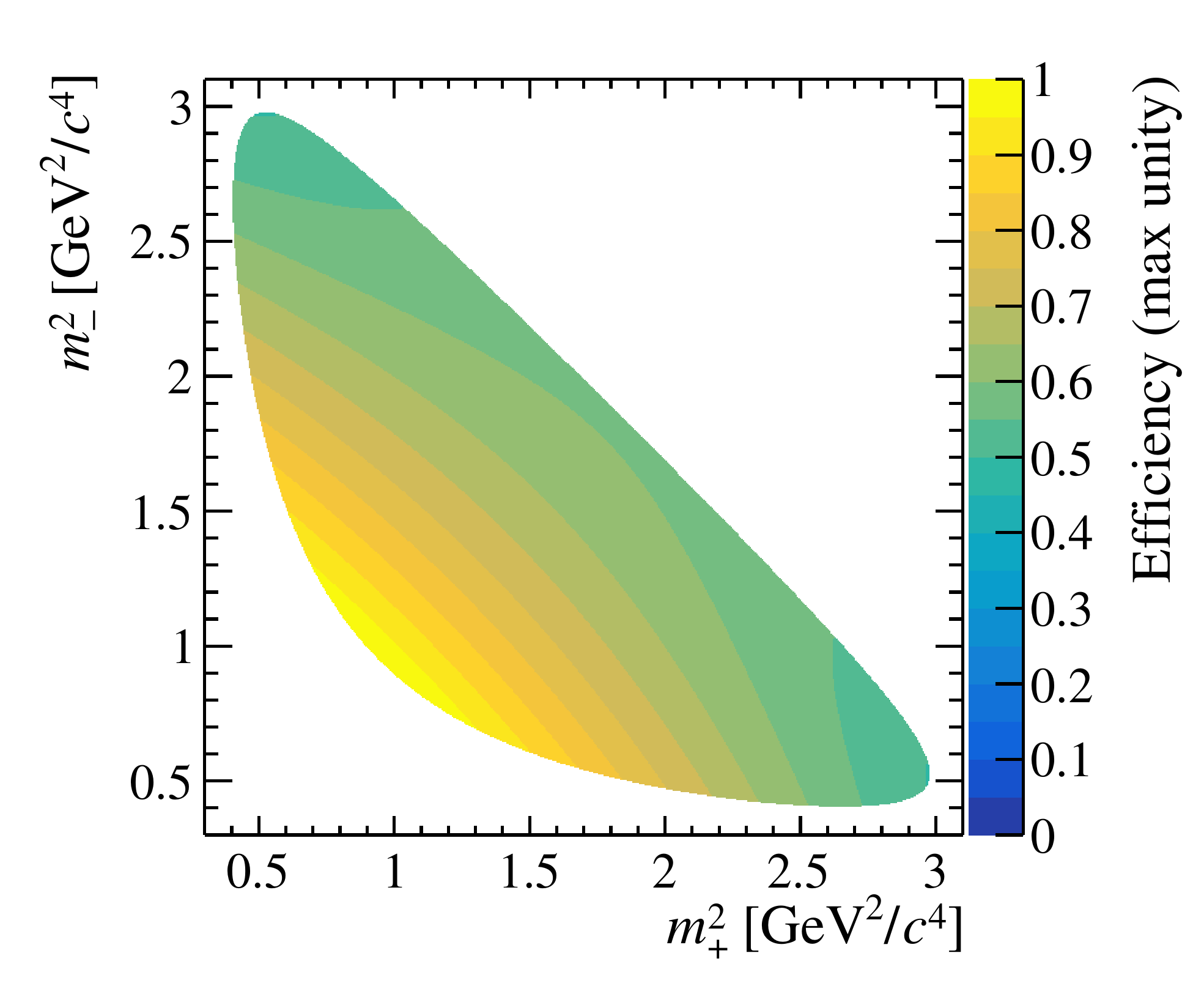}\\
\includegraphics[width=0.6\textwidth]{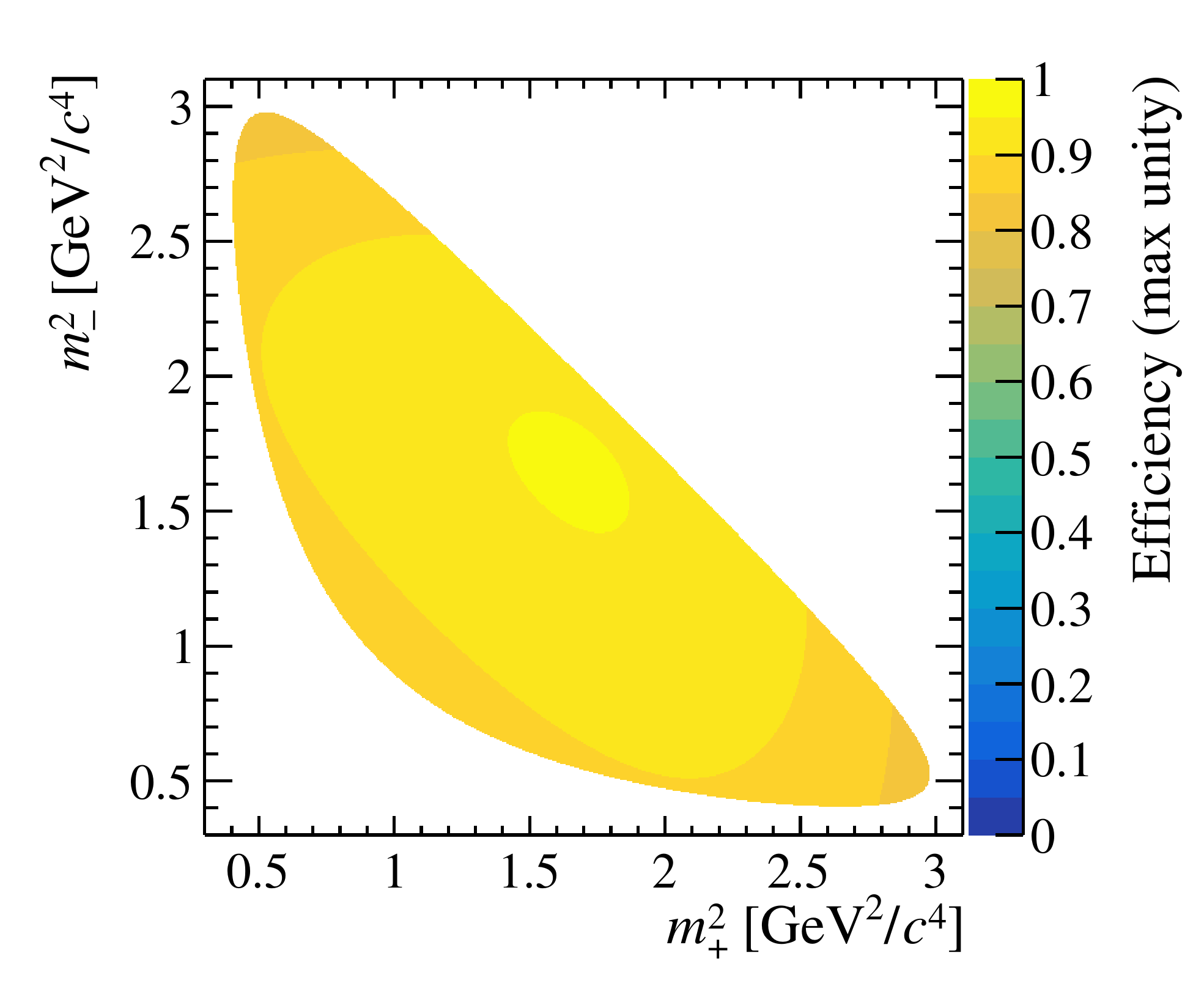}\\
\caption{Efficiency (normalized to unity at its maximum) as a function of the Dalitz plot location assumed for \mbox{\Dkspp} decays reconstructed from the $\Dstar(2010)^+\to\Dz\pip$ decay chain with the (top) LHCb and (bottom) Belle~II detectors.\label{fig:dalitz_acc}}
\end{figure}

For LHCb, we assume a decay-time resolution corresponding to 10\% of the \Dz lifetime, the decay-time-dependent variation of the reconstruction efficiency shown in \cref{fig:prompt_dt_acc}, and the efficiency variation over the Dalitz plane of \cref{fig:dalitz_acc}, following Refs.~\cite{Pilar:2014jfg,Torr:2004384}. For Belle~II, we assume a decay-time resolution corresponding to 33\% of the \Dz lifetime following Ref.~\cite{Kou:2018nap}. For convenience, we use the model of the efficiency variation over the Dalitz plot derived for the BaBar detector, in \cref{fig:dalitz_acc}~\cite{GarraTico:2010zz}. This is unlikely to be an accurate model for the Belle~II efficiency, but it is sufficient for the purpose of demonstrating the robustness of the method against nonuniformities in efficiency, which ought not depend on the details of the efficiency itself. We also assume that the Belle~II reconstruction efficiency is uniform as a function of decay time. For both LHCb and Belle~II, the resolutions on $m_+^2$ and $m_-^2$ are similar to or better than the $0.0054\,\massgevsq$ spacing used by CLEO to define the Dalitz bins. Because such spacing is significantly smaller than the typical size of the Dalitz bins, $m_+^2$ and $m_-^2$ resolutions are expected to introduce negligible bin migrations and are therefore neglected.

Data are generated using the same amplitude model as for the previous studies. The decay-time resolution is included by smearing the generated decay time with a Gaussian distribution with a width of $0.1\tau$ ($0.33\tau$) for the samples simulating LHCb (Belle~II) conditions. The effects of selection requirements on the decay-time and Dalitz-plot distributions are incorporated by sampling the generated events according to the relevant parametrizations. The analysis procedure is then repeated as previously described, without modeling the resolution and efficiency effects in the fits.

\begin{table}[t]
\centering
\caption{Biases ($B$), normalized to the statistical uncertainty ($\sigma$), due to neglecting efficiency and resolution effects expected at LHCb and Belle~II, as functions of the number of events. We use simulated \Dkspp decays generated in the WM scenario, and fit allowing $r_b$ to float freely and keeping $(c_b,s_b)$ constrained. The constraint assumes the current (improved) determination of the external measurements of $(c_b,s_b)$ for 1--$10\times10^6$ (5--$50\times10^7$) signal yields.\label{tab:detector_results}}
\begin{tabular}{lrrrr}
\toprule
Signal yield & $B/\sigma(\xcp)$ & $B/\sigma(\ycp)$ & $B/\sigma(\deltax)$ & $B/\sigma(\deltay)$ \\
\midrule
LHCb detector \\
$\quad1\times10^6$ & $0.04$ & $0.02$ & $0.05$ & $0.02$ \\
$\quad5\times10^6$ & $0.07$ & $0.07$ & $0.03$ & $0.10$ \\
$\quad1\times10^7$ & $0.03$ & $0.09$ & $0.10$ & $0.09$ \\
$\quad5\times10^7$ & $0.10$ & $0.05$ & $0.27$ & $0.15$ \\
$\quad1\times10^8$ & $0.12$ & $0.09$ & $0.40$ & $0.16$ \\
$\quad5\times10^8$ & $0.22$ & $0.10$ & $1.00$ & $0.42$ \\
\midrule
Belle~II detector \\
$\quad1\times10^6$ & $0.03$ & $0.06$ & $0.07$ & $0.05$ \\
$\quad5\times10^6$ & $0.17$ & $0.06$ & $0.08$ & $0.06$ \\
$\quad1\times10^7$ & $0.18$ & $0.03$ & $0.05$ & $0.03$ \\
$\quad5\times10^7$ & $0.40$ & $0.07$ & $0.04$ & $0.03$ \\
\bottomrule
\end{tabular}
\end{table}

The fits are performed with unconstrained $r_b$ parameters and $(c_b,s_b)$ parameters constrained. The constraint assumes that the precision of the external measurements of $(c_b,s_b)$ is improved by a factor two (four) at sample sizes of $5\times10^7$ (1--$5\times10^8$) signal decays. \Cref{tab:detector_results} lists the magnitudes of the biases with respect to the generated values, normalized to the fit uncertainties, as functions of sample size. In the LHCb case, the observed biases are mostly due to neglecting efficiency variations across the Dalitz plane. For Belle~II, neglecting the decay-time resolution dominates. As expected, the relative impact of small constant biases becomes more significant as the statistical precision of the measurements increases. The largest effect is observed for LHCb, with a \deltax bias comparable with the statistical uncertainty in the highest signal-yield scenario. All other biases do not exceed $40\%$ of the statistical uncertainty. 

These findings show that no accurate knowledge of the decay-time resolution or efficiency variation as a function of decay time and Dalitz-plane position is needed to apply the method. This supports the approach as an expedient and powerful alternative to standard approaches for charm-mixing measurements using \mbox{\Dkspp} and other multibody decays in current and next generation analyses. Further refinements will probably be needed to fully exploit the method at the very high yields expected a decade from now in the final LHCb sample. 

\section{Impact on knowledge of charm-mixing parameters\label{sec:impact}}

\begin{figure}[th]
\centering
\includegraphics[width=0.5\textwidth]{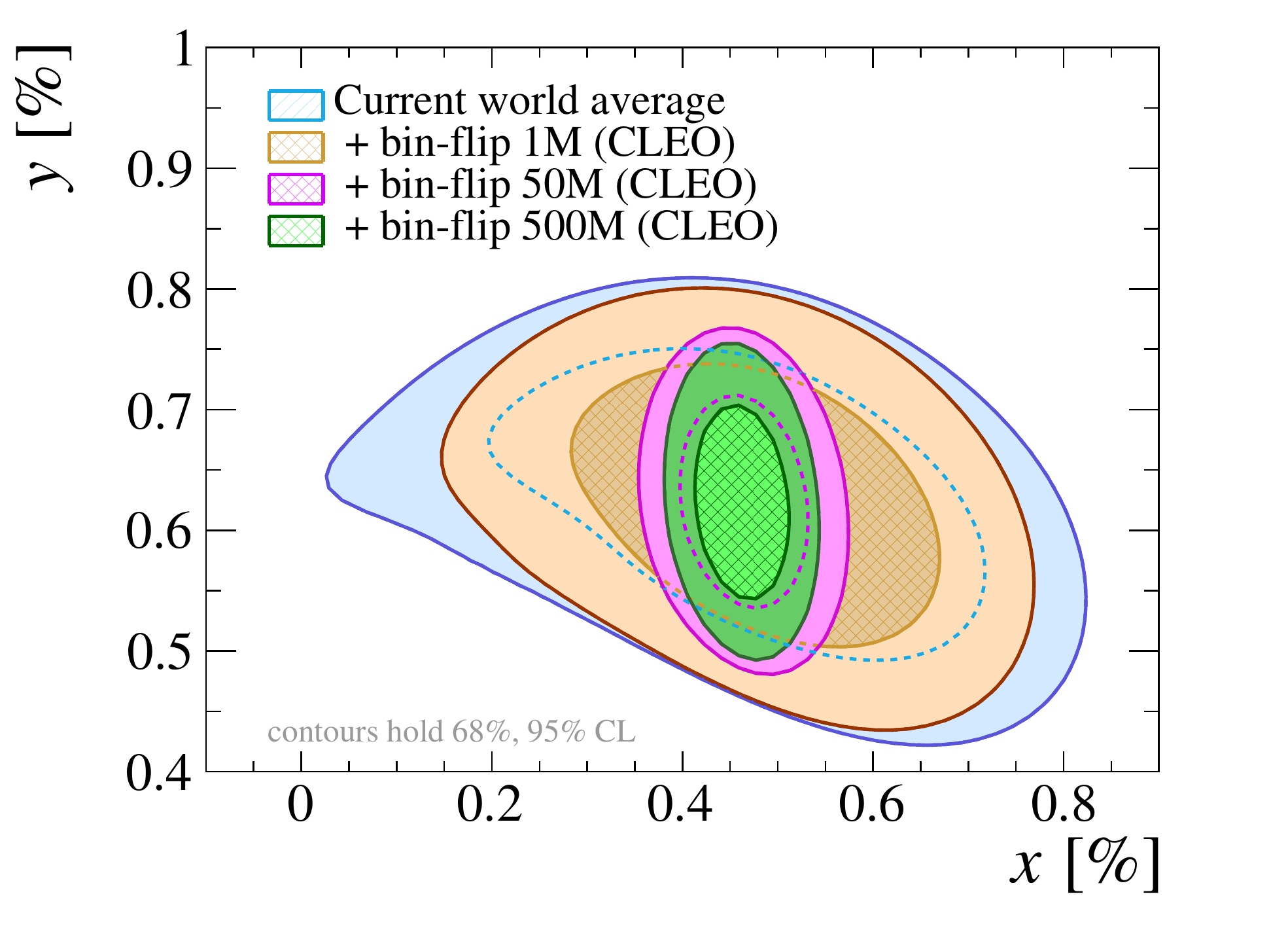}\hfil
\includegraphics[width=0.5\textwidth]{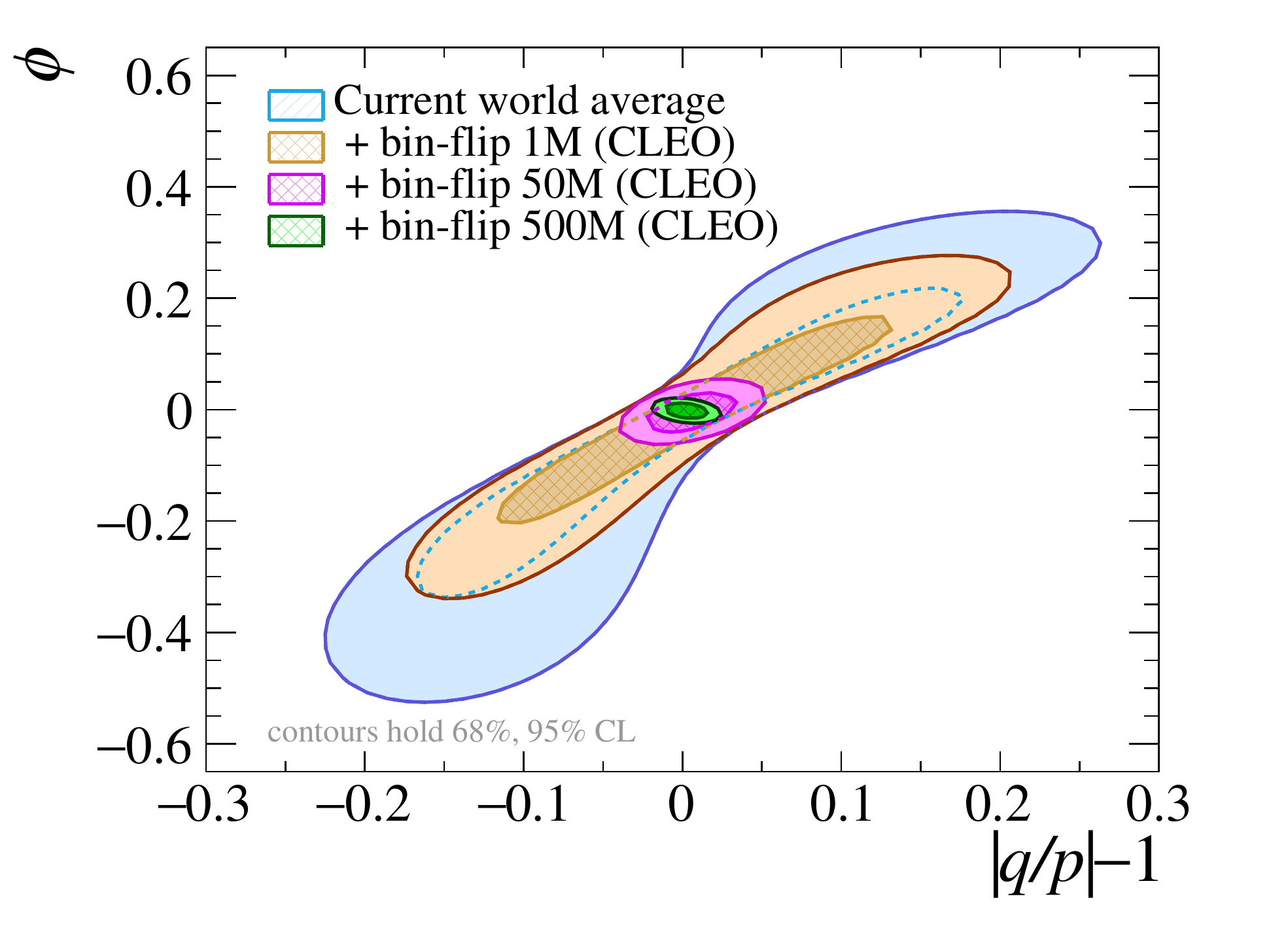}\\
\includegraphics[width=0.5\textwidth]{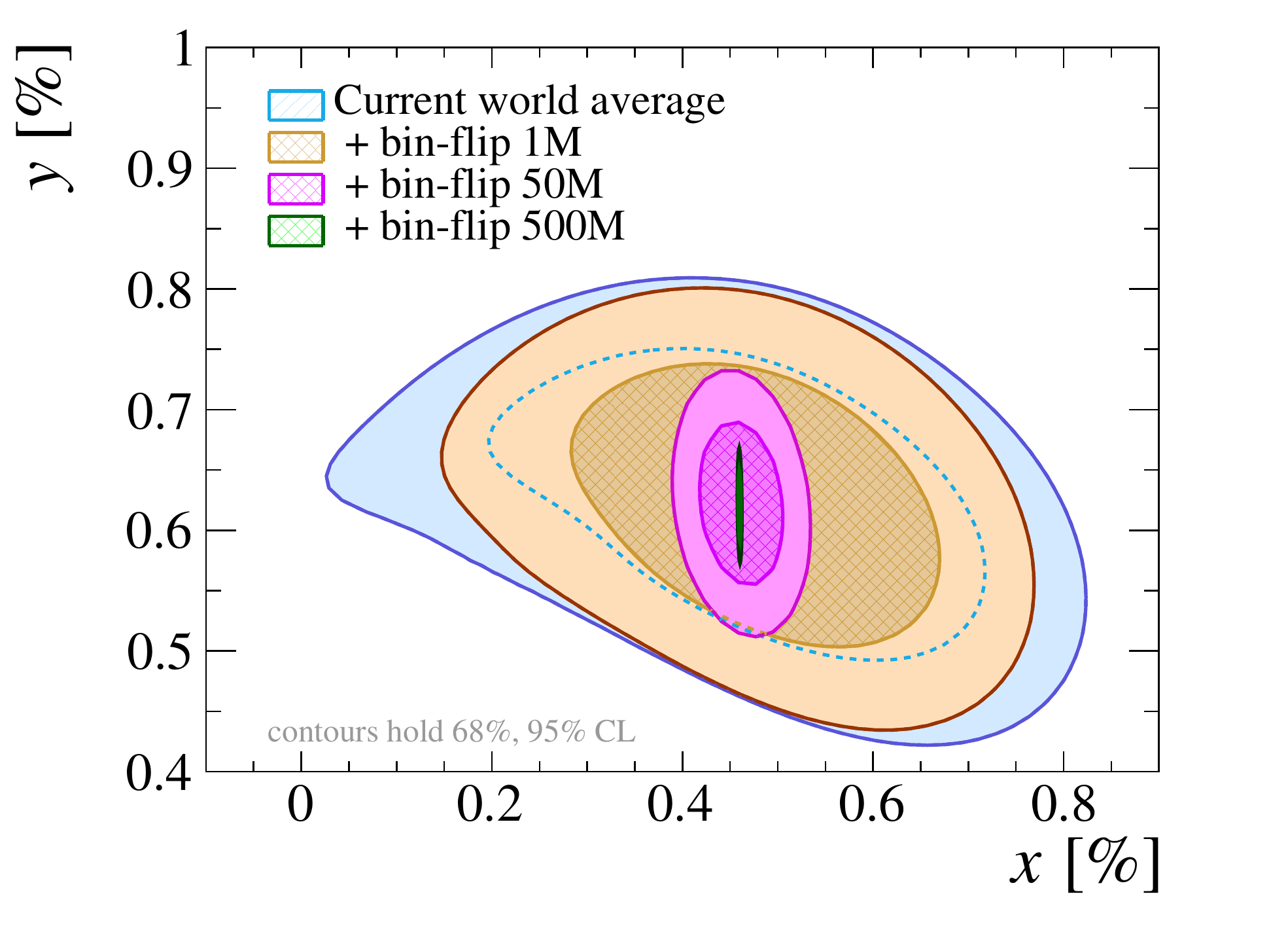}\hfil
\includegraphics[width=0.5\textwidth]{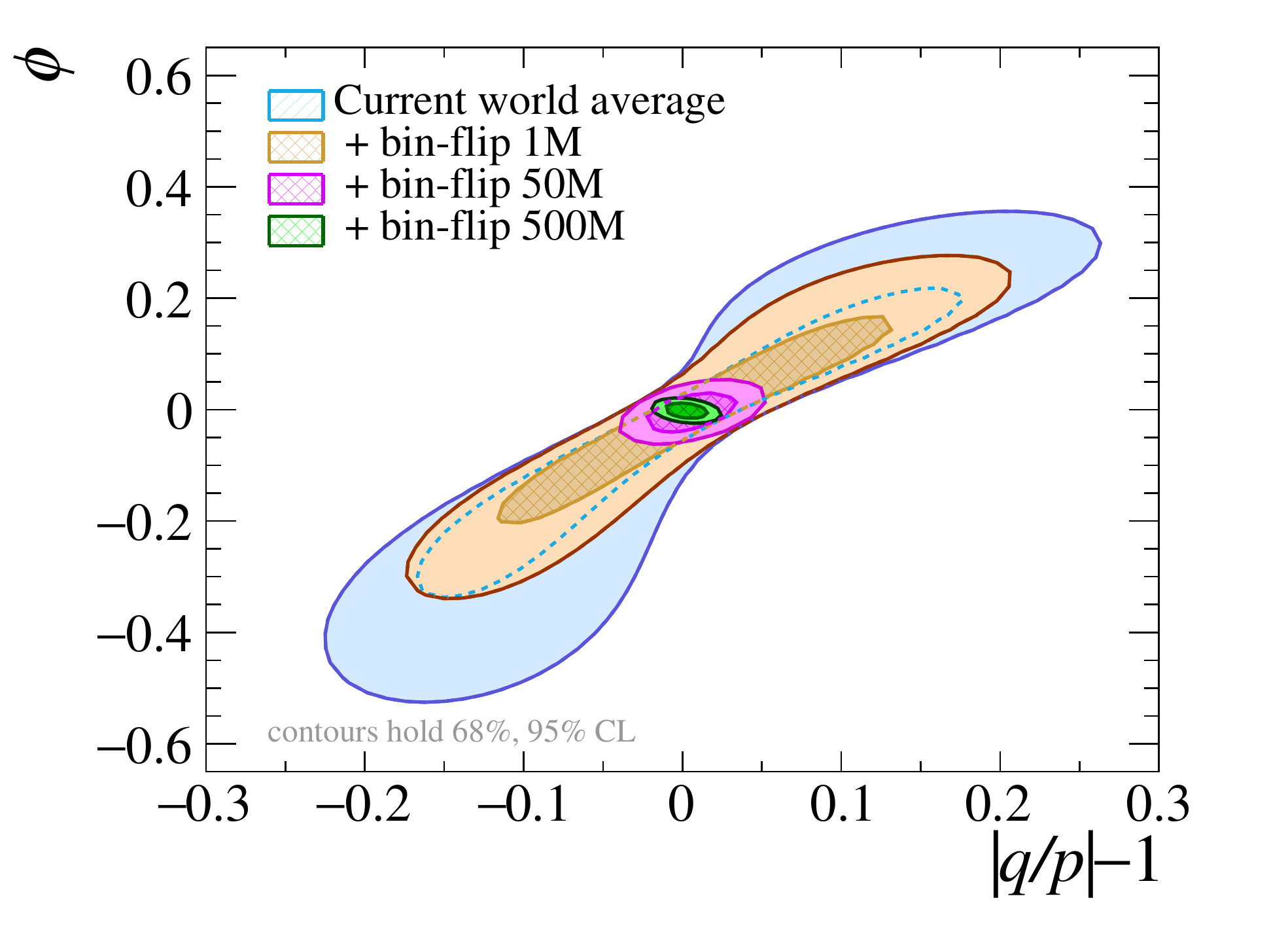}\\
\caption{Confidence regions at the (inner, darker hatching) 68.3\% and (outer, lighter hatching) 95.5\% confidence level in the two-dimensional space of (left) oscillation parameters ($x$, $y$) and (right) parameters of \CP violation in mixing and in the interference between mixing and decay ($|q/p|-1$, $\phi$) corresponding to (blue) current world-average results and to those results updated to include a bin-flip analysis of (orange) $1\times10^6$, (magenta) $5\times10^7$, and (green) $5\times10^8$ signal decays. Top panels refer to results based on current CLEO inputs on $(c_b,s_b)$ parameters; bottom panels on improved $(c_b,s_b)$ inputs. The displayed regions assume unchanged central values and precision of bin-flip results dominated by statistical uncertainties.\label{fig:impact}}
\end{figure}

To assess the impact of a bin-flip analysis on the current global knowledge of oscillation and \CP-violation parameters in charm mixing, we compare the precision of the current world-average determination of $x$, $y$, $\phi$, and $|q/p|$, with the precision achievable when including a bin-flip analysis of $1\times10^6$, $5\times10^7$, and $5\times10^8$ \Dkspp decays. \Cref{fig:impact} shows the results assuming unchanged central values, precision of bin-flip results dominated by statistical uncertainties, and either current or improved determination of the external measurements of $(c_b,s_b)$ parameters.

While the effect on $y$ is relatively minor, the bin-flip method is expected to have a major impact in the determination of $x$ and of the \CP-violation parameters. For instance, the comparison between the current world-average constraints (blue region), with their update including bin-flip results based on $10^6$ signal decays (orange region), offers a realistic representation of the impact the bin-flip analysis could have if applied to typical current LHCb samples. Consistently with \cref{tab:gen-toy-uncertainties-vs-n}, the precision of the external inputs has negligible impact on the determination of the \CP-violation parameters $|q/p|$ and $\phi$ but will strongly enhance the reach in $x$ and $y$ when larger samples will be analyzed.

We finally emphasize that the alternative additive parametrization proposed for the effects of charm mixing offers superior statistical properties to standard parametrizations and is particularly preferable for combinations, in which central values cannot be assumed to be known.

\section{Conclusions\label{sec:conclusions}}

In summary, we propose the bin-flip method, a model-independent approach to measure parameters of mixing and \CP violation in charm from multibody decays, such as \mbox{\Dkspp}. The method avoids the need for accurate knowledge of either the decay-time resolution or the signal-reconstruction efficiency as a function of decay time and position in the Dalitz plot. We also introduce a novel parametrization of the oscillation and \CP-violation effects in charm mixing that has attractive statistical properties and may find wider applicability.

The bin-flip method offers 35\% better statistical sensitivity, compared to existing model-independent methods, to \CP-averaged and \CP-violating quantities related to the mass difference between the neutral \D eigenstates, while suppressing systematic effects due to nonuniform efficiencies in decay time and across the Dalitz plane. In addition to the gain in precision, the demonstrated insensitivity to the details of Dalitz-plot and decay-time modeling make the application of this method significantly simpler and more expedient than other model-independent approaches, especially in hadron-collision experiments.

The bin-flip method is expected to offer good sensitivity in high-yield multibody decays that receive large contributions from doubly Cabibbo-suppressed amplitudes. In addition to \Dkspp decays, it is likely to benefit the analysis of $\Dz\to\KS\pip\pim\piz$ decays, for which first measurements of the relevant hadronic parameters have recently become available~\cite{K:2017qxf}. The bin-flip method can also, with straightforward modifications to the formalism, be used with decays to non-self-conjugate final states such as $\Dz \to \Kmp\pipm\piz$ and $\Dz\to\Kmp\pipm\pip\pim$. Conversely, the sensitivity is reduced in channels where \CP-eigenstate amplitudes dominate in many of the Dalitz-plot bins, such as $\Dz\to\KS\Kp\Km$, $\Dz\to\pip\pim\piz$ and $\Dz\to \Kp\Km\piz$~\cite{Gershon:2015xra}.

A bin-flip analysis of the samples of \Dkspp decays expected to be collected at the LHCb or Belle~II experiments has the potential to significantly improve the global knowledge of the charm-mixing parameters and yield more stringent constraints on \CP violation in charm oscillations. The method is expected to avoid limiting systematic uncertainties even with very large data samples, when improved knowledge of the hadronic $(c_b,s_b)$ parameters from independent measurements will help to achieve even better precision.

\section*{Acknowledgements\addcontentsline{toc}{section}{Acknowledgements}}
We are grateful to Andrea Contu and Michal Kreps for earlier involvement in this work, and Jolanta Brodzicka for fruitful discussions and valuable comments. TG and NJ acknowledge support from the Science and Technology Facilities Council (United Kingdom). TG and TP acknowledge support from the European Research Council under FP7.

\setboolean{inbibliography}{true}
\addcontentsline{toc}{section}{References}
\bibliographystyle{LHCb}
\bibliography{main,LHCb-PAPER}
\setboolean{inbibliography}{false}

\cleardoublepage
\appendix
\section{\boldmath An alternative parametrization of \CP violation in charm mixing\label{app:cpvparametrization}}

In \cref{sec:method}, we introduced a new parametrization of charm-mixing effects expressed as functions of the additive parameters \zcp and \deltaz, defined by
\begin{equation}
\zcp\pm\deltaz\equiv\left(q/p\right)^{\pm1}z,
\end{equation}
in terms  of the conventional multiplicative parameters $z$ and $q/p$. The proposed parametrization offers nontrivial advantages in the determination of parameters from fits to data.

\begin{figure}[h]
\centering
\includegraphics[width=0.5\textwidth]{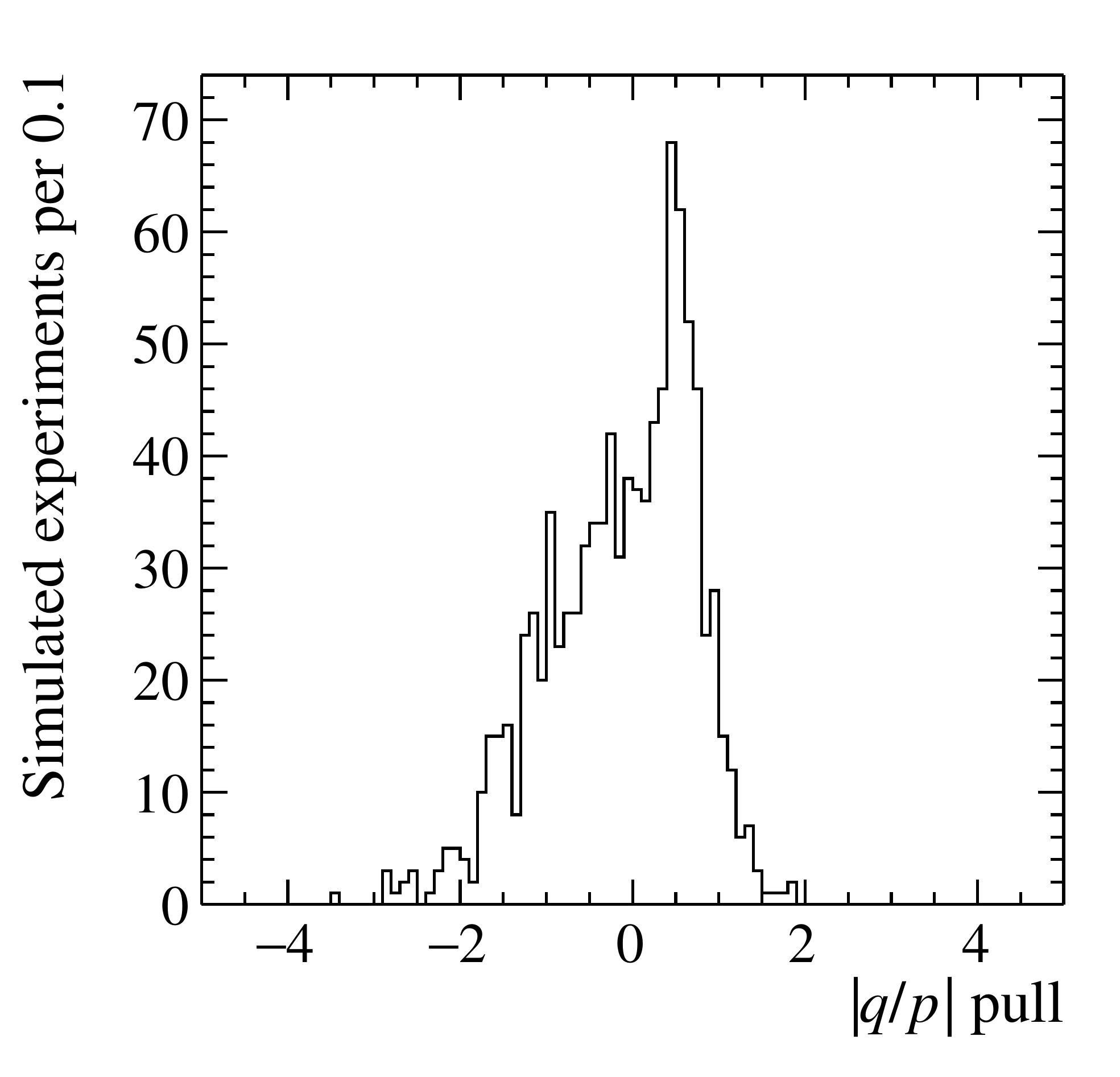}\hfil
\includegraphics[width=0.5\textwidth]{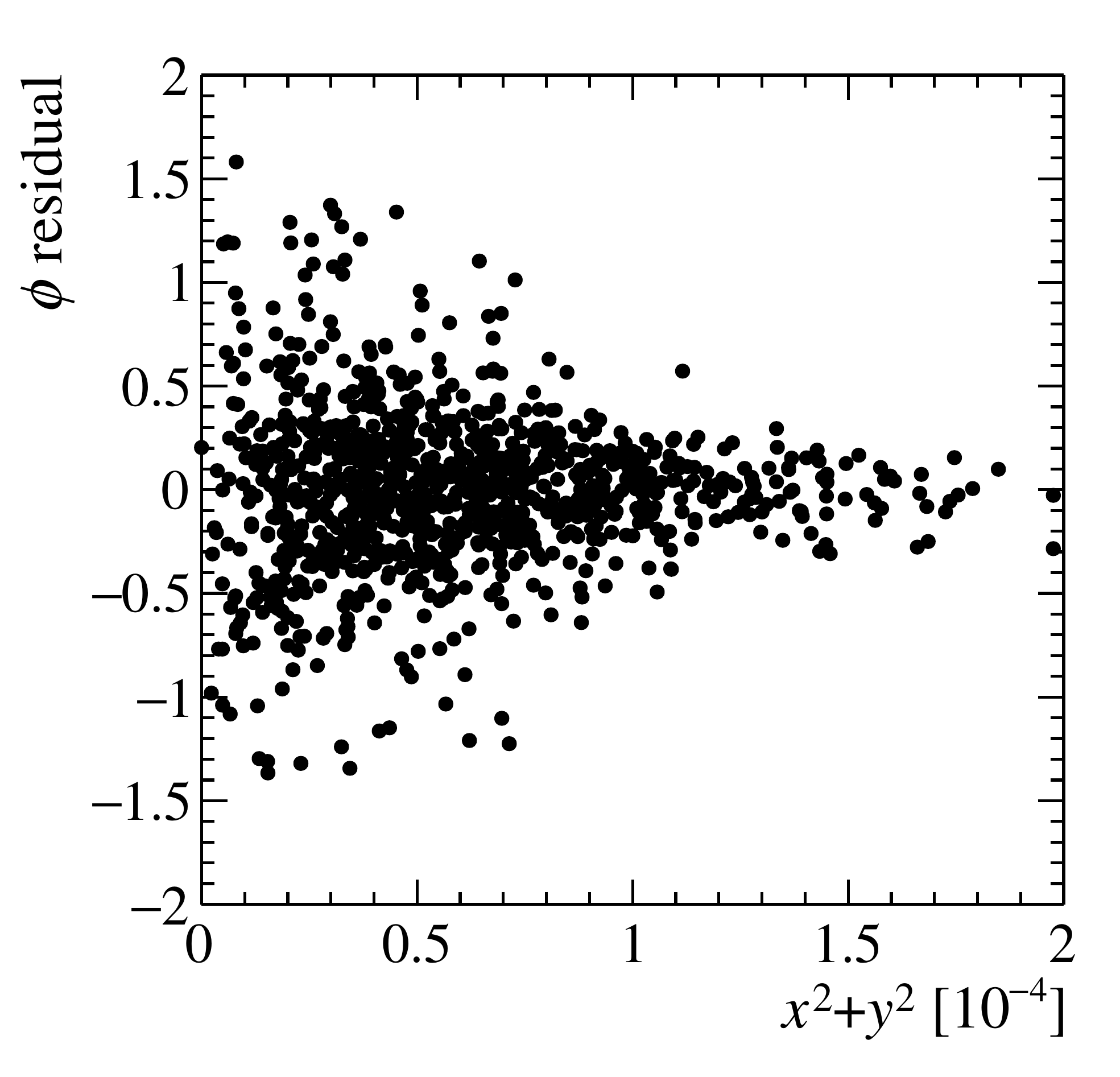}\\
\caption{Distributions of (left) fit pull for the $|q/p|$ parameter and (right) fit residual for the $\phi$ parameter as functions of the observed mixing rate $x^2+y^2$.\label{fig:parametrization-issues}}
\end{figure}

Fits suffer from non-Gaussian estimator distributions when the dimensionality of the likelihood or least-squares function depends on the estimated value of one or more parameters. This may happen if all terms sensitive to a parameter of interest involve products with another parameter, or a function of it, that can vanish. The likelihood then becomes scarcely sensitive to the parameter of interest for vanishing values of the multiplication factor, incurring in non-Gaussian estimator distributions.  A multiplicative parametrization as $(q/p)^{\pm1}z$ is prone to such effects, as shown using simulated experiments in the WM scenario in \cref{fig:parametrization-issues}: $|q/p|$ pulls are non-Gaussian and the dispersion of the $\phi$ residual depends on the observed mixing rate. These issues are avoided when using our parametrization in terms of \zcp and \deltaz, as shown in \cref{sec:sensitivity}.

\end{document}